\documentclass[manuscript]{acmart}

\usepackage[utf8]{inputenc}
\usepackage{xcolor}
\usepackage{booktabs}
\usepackage{graphicx}
\usepackage{multirow}
\usepackage{soul}
\usepackage{fancybox}
\usepackage{comment}
\usepackage{subcaption}
\usepackage[title]{appendix}

\newcounter{observation}
\cornersize{.2}
\newcommand{\observation}[1]{\refstepcounter{observation}
	\begin{center}
		\Ovalbox{
			\begin{minipage}{0.93\columnwidth}
				{\bf Observation \arabic{observation}:} #1
			\end{minipage}
		}
	\end{center}
}

\title{Finding the Needle in a Haystack: Detecting Bug Occurrences in Gameplay Videos}
\author{Andrew Truelove}
\affiliation{%
    \institution{Donald Bren School of Information and Computer Sciences, University of California, Irvine}
    \streetaddress{Donald Bren Hall}
    \city{Irvine}
    \state{CA}
    \country{USA}
    \postcode{92697-3425}
}
\email{truelova@uci.edu}
\author{Shiyue Rong}
\affiliation{
    \institution{Donald Bren School of Information and Computer Sciences, University of California, Irvine}
    \streetaddress{Donald Bren Hall}
    \city{Irvine}
    \state{CA}
    \country{USA}
    \postcode{92697-3425}
}
\email{shiyuer@uci.edu}
\author{Eduardo Santana de Almeida}
\affiliation{%
    \institution{Institute of Computing, Federal University of Bahia (IC-UFBA)}
    \streetaddress{Avenida Milton Santos, s/n - Ondina Campus, PAF 2}
    \city{Salvador}
    \state{Bahia}
    \country{Brazil}
    \postcode{40.170-110}
}
\email{esa@rise.com.br}
\author{Iftekhar Ahmed}
\affiliation{%
    \institution{Donald Bren School of Information and Computer Sciences, University of California, Irvine}
    \streetaddress{2438 ISEB}
    \city{Irvine}
    \state{CA}
    \country{USA}
    \postcode{92697-3440}
    }
\email{iftekha@uci.edu}

\date{}

\begin{document}

\begin{abstract}

The presence of bugs in video games can bring significant consequences for developers. To avoid these consequences, developers can
leverage gameplay videos to identify and fix these bugs. Video hosting websites such as YouTube provide access to millions of game
videos, including videos that depict bug occurrences, but the large amount of content can make finding bug instances challenging. We present an automated
approach that uses machine learning to predict whether a segment of a gameplay video contains the depiction of a bug. We analyzed
4,412 segments of 198 gameplay videos to predict whether a segment contains an instance of a bug. Additionally,
we investigated how our approach performs when applied across different specific genres of video games and on
videos from the same game. We also analyzed the videos in the dataset to investigate what characteristics of the visual
features might explain the classifier’s prediction. Finally, we conducted a user study to examine the benefits of
our automated approach against a manual analysis. Our findings indicate that our approach is effective at detecting segments
of a video that contain bugs, achieving a high F1 score of 0.88, outperforming the current state-of-the-art technique for bug classification of gameplay video segments.

\end{abstract}

\maketitle

%!TEX root = main.tex

\section{Introduction}
\label{sec:intro}
 
The video game industry is a multi-billion dollar industry, expected to exceed \$211 billion in revenue by 2025~\cite{gamesMarket_2022_Newzoo}.
Illustrating this fact, in 2018, the video game \textit{Grand Theft Auto V} was the most profitable media title of all time, having earned an estimated \$6 billion in revenue~\cite{cherney_this_2018M}.
Due to such financial impact, presence of bugs in these video games are consequential for developers. For example, when development studio CD Projekt Red released \textit{Cyberpunk 2077} in December of 2020, the prevalence of bugs led to a widespread negative reception~\cite{carpenter_cd_2020M}, the removal of the game from digital storefronts~\cite{dornbush_sony_2020M}, and a 41\% drop in the company's stock price~\cite{isaac_cyberpunk_2020M}.

To avoid these kinds of negative impacts, it has become common practice for game developers to apply updates to their games in order to fix bugs after the game is released; for example, over one third of the big-budget games released in 2014 received a day-one update~\cite{NarcisseDayOnePatch}.  
This is because games can be complex, with ``enormous alternative ways that players can execute the software, differently from individuals using a software with a well-defined number of actions''~\cite{santos_computer_2018}; this complexity can lead to difficulties in thoroughly testing for all possible bugs. 
As such, game developers have become increasingly interested in leveraging gameplay videos to help identify bugs that have persisted into a game's release~\cite{lin_identifying_2019}. 
Gameplay videos are useful for developers, because they offer a wide range of information such as the state of the game at the time a bug appeared, the audio content at the time of the bug, the visual content at the time of the bug, etc.~\cite{lin_identifying_2019}.  
Gameplay videos are often easily accessible as well; video hosting services such as YouTube~\cite{youTubeM} and Twitch~\cite{twitchM} allow users to upload, stream, and view gameplay videos, including videos depicting the occurrences of bugs~\cite{lin_identifying_2019}. Gaming videos are particularly popular on these sites; 2020, for example, saw the number of active gaming channels on YouTube surpass 40 million~\cite{watt_2020_2020M}.
While these videos can provide developers with useful information about bugs, the large amount of video content on sites like YouTube create difficulties in locating relevant footage of bug incidents~\cite{lin_identifying_2019}. 
Aside from the quantity of videos available on the platform, there is also the issue of individual video length. For example, the most liked non-live gaming video on YouTube in 2020 was over 18 minutes long~\cite{watt_2020_2020M, fernanfloo_i_2020M}, and prior research on the Twitch video platform found that gameplay videos from popular streamers frequently lasted between ``two and eight hours in length''~\cite{sjoblom2019ingredients}. In cases like these, it is not enough to know that a video contains a bug. Rather, a developer needs to know when the bug actually occurs. Otherwise, they would need to watch the video in its entire duration to locate the bug, which becomes burdensome when multiple videos with long lengths are involved. 

In this paper, we propose a novel automated approach to split a video of gameplay into smaller segments and predict which segments depict the occurrence of bugs. 
%We implement a feature-extraction scheme that involves clustering frame images from the videos in a dataset, designating representative frames for each cluster, and comparing these key frames with the key frames from each of the other clusters.
We also perform a preliminary analysis to explain which characteristics of the video frame images are more likely to have an association with the occurrence of bugs.
Our approach improves upon prior work by Guglielmi et al., which focused on devising an approach to identify segments of YouTube videos that depict anomalies in gameplay videos and classify the segment based on whether it contained the presence of a game bug~\cite{guglielmi2023using}. While their approach relied on the presence of certain keywords in the video's transcript to locate relevant buggy segments, our approach is able to identify and classify segments regardless of the content of the transcript text. Additionally, our classifier outperforms their best-performing classifier configuration when run on our dataset of video segments.

In addition to the classification task, we also perform the novel task of conducting an analysis on frames from the bug-depicting segments in our dataset to identify characteristics of gameplay videos that appear to have an association with bugs. Developers are more likely to use a machine learning tool if they feel they can trust the model, and the ability to explain a model's predictions is an important aspect of establishing that trust~\cite{ribeiro_why_2016}. Additionally, determining specific objects or behaviors from these video segments can inform developers on what kinds of game scenarios may be particularly bug-prone. This knowledge can be used to investigate additional videos that depict similar material and to bolster testing to better target these scenarios. In our analysis, we wanted to identify the elements from the video frame that a model believed to be important in making a prediction, the relationship between those highlighted elements and the depicted bug, and the characteristics and patterns that the model seemed to pick up on more frequently.     

For our study, we selected 198 buggy game videos from YouTube and automatically split them into 4,412 video segments. We extracted textual transcript information and visual information from these segments for use as features, as these features have been used for video classification tasks in prior research~\cite{brezeale2008automatic}.    
We train a collection of neural network-based classification models with the aim of predicting whether a video segment shows the occurrence of a bug. Additionally, we evaluated the models on subsets of our dataset containing only video segments from games of the same genre as well as datasets where all segments were from the same game. 
We also run the technique by Guglielmi et al. on all our datasets as a baseline and compare its performance with the performance of our classifiers. %evaluate the performance of our models against the performance of the technique by Guglielmi et al.. 
We then analyze video frames in our dataset using LIME superpixel explanations~\cite{ribeiro_why_2016} in order to identify characteristics of game images that can help explain the classifier's prediction. Finally, we conduct a user study to evaluate the benefits of our framework when compared to manual analysis.

Our research questions are as follows:

%\textcolor{blue}{
\textbf{RQ1}: How well does our tool perform in predicting bug occurrences in gameplay videos? 

%\textcolor{blue}{
\textbf{RQ2}: How does the tool perform when trained on footage from different genres of video games?

%\textcolor{blue}{
\textbf{RQ3}: How does the tool perform while predicting bug occurrences across games? 

\textbf{RQ4}: Is it possible to identify which characteristics of visual video frames have an association with the prediction of a bug?

\textbf{RQ5}: What benefits are brought to developers using our tool in identifying buggy gameplay segments in videos over a manual analysis?

The rest of our work is structured as follows.
In Section~\ref{sec:related}, we review related works that are similar to our research problem.
In Section~\ref{sec:method}, we describe the process of developing our automated bug identification approach approach, including our data preparation procedure (\ref{subsec:dataPrep}), our classification models used with the different training sets (\ref{subsec:ClassModels}), our superpixel explanations (\ref{subsec:explaining}), our video attributes (\ref{subsec: videoAttributes}), and our user study (\ref{sebsec:userStudy}). Then, in Section~\ref{sec:results}, we present the results to our research questions, including the evaluation of our different bug identification models.
Section~\ref{sec:discussion} provides a discussion of our findings, including implications for both researchers and practitioners. We go over threats to validity in Section~\ref{sec:threats}, and finally, we present our conclusions in Section~\ref{sec:conclusion}.

Our contributions in the paper are as follows: 
\begin{itemize}
\item An automated tool that uses machine learning to predict whether a segment of a gameplay video contains the depiction of a bug through the sole use of features from the video itself.
\item A dataset comprised of 4,412 gameplay video segments from 198 different game videos.
\item An analysis of characteristics from the gameplay videos that help explain which characteristics are likely to have an association with bugs.
\end{itemize} 
%!TEX root = main.tex

\section{Related Work}
\label{sec:related}

%\textbf{
Prior work has focused on analyzing game videos in order to extract particular information, including information relating to bugs. %} 
For example, 
Lin et al. developed a method to identify game play videos that showcase bugs; their method involved an analysis of metadata for videos on both YouTube and Steam~\cite{lin_identifying_2019}. Their method was was able to achieve precisions of over 0.90 when evaluated on a set of 1,400 known buggy videos~\cite{lin_identifying_2019}. 
Lewis et al. created a taxonomy for video game bugs by examining online game play videos and reading online articles related to game bugs~\cite{lewis_what_2010}.
Amiriparian et al. developed a technique to analyze game audio from YouTube game videos with the purpose of identifying the genre of the game being played~\cite{amiriparian_are_2019}. An evaluation of their method returned ``an accuracy of up to 66.9\% unweighted average
recall using tenfold cross-validation''~\cite{amiriparian_are_2019}.
Shah et al. developed an approach for training a convolutional neural network model to generate automated commentary from a given input video~\cite{shah_automated_2019}. They trained the model using features from YouTube videos--specifically, for each video they extracted the frame images from the video and the transcript of the video's commentary~\cite{shah_automated_2019}. 
Taesiri et al. developed a method for querying videos of game bugs in order to match them with a relevant gameplay video from a dataset~\cite{taesiri2022clip}. The approach involved taking a textual input query describing a particular object or event; the approach then identified image frames in the dataset with a high similarity score with the query, returning the videos that were most likely to depict they subject of the query~\cite{taesiri2022clip}. The authors' proposed this technique could be used to search for and identify gameplay videos that depict certain types of bugs (i.e., a query asking for ``a person stuck in a horse' returns a video depicting that specific type of bug)~\cite{taesiri2022clip}.
Guglielmi et al. introduced GELID, an autommated approach to extract segments from gameplay videos related to bugs~\cite{guglielmi2023using}. To approach is designed to return meaningful video information by (i) extracting relevant segments of a video likely to contain a bug, (ii) automatically classifying the segments as informative or non-informative with respect to the presence of a bug (as well as, in the informative case, the type of bug being depicted), and (iii) groups the segments based on the context in which the bug appears (e.g., the game location or level), as well as (iv) grouping based on the type of bug being depicted~\cite{guglielmi2023using}. While GELID was able to identify informative segments, their classifier was only able to achieve an average F1 score of 0.63 for binary bug classification across their test datsets, which the authors concluded were not satisfactory results~\cite{guglielmi2023using}. 

Our work is distinguishable from these works in the sense that none of these works seek to identify the timestamps at which bugs appear in gameplay videos. 
Amiriparian et al. and Shah et al. were interested in problems other than game bugs~\cite{amiriparian_are_2019, shah_automated_2019}. 
Lewis et al. looked at the types of bugs that appeared in gameplay videos, not when they appeared~\cite{lewis_what_2010}. 
Taesiri et al. focused on returning videos that depicted the objects or events requested in the query, but their approach did not return information about when exactly in the video the queried information appeared~\cite{taesiri2022clip}. Additionally, the approach was specifically evaluated on a specific type of bug--'game physics' bugs--and its effectiveness on identifying other types of bugs was not examined~\cite{taesiri2022clip}. Our approach is not strictly intended with a particular type of bug in mind. 
While Lin et al. were also engaged in identifying the presence of bugs in game videos, their identification method primarily determines whether a video showcases a bug and does not provide any time specification information~\cite{lin_identifying_2019}. A two-minute video and a two-hour video could return the same positive result, requiring a manual analysis of the video to identify when the bug appears~\cite{lin_identifying_2019}. 
Additionally, the features used for their classifier were extracted from video metadata, which included the video title, description, and tags~\cite{lin_identifying_2019}. 
While this approach can help developers identify videos that might showcase a bug, the method is lacking when that metadata is absent or unrelated to bugs, such as in the case of a video in which a bug appears in gameplay, but it is not necessarily the main focus of the video. By contrast, our approach used features directly from the video itself, meaning that extraction of additional metadata is not necessary.

The work by Guglielmi et al. is likely the closest work to ours, since it is also focused on automatically identifying bugs from segments of gameplay videos as one of its steps~\cite{guglielmi2023using}. Our work can be distinguished in a few ways, however. First, the process they used for their classification involved discarding or excluding segments that did not contain certain keywords in their transcripts from their data~\cite{guglielmi2023using}. As such, their approach may not be geared towards handling segments lacking these keywords. On the other hand, our approach is designed so that it can process video segments regardless of the contents of the transcript text. Additionally, our work can be distinguished in the sense that we focused on using different models trained on specific subsets of video game segments, including a general dataset of segments, datasets containing only segments from the same game genre, and segments from the same individual game. While Guglielmi et al. tested their models on three datasets with each set containing segments from a single game, this generalized and genre-level analysis are not considered~\cite{guglielmi2023using}. Finally, the authors acknowledged that the performance of the binary GELID classifier was not satisfactory, achieving an average F1 score of only 0.63 for the positive class across the three game-level test sets and a top F1 score of 0.68~\cite{guglielmi2023using}. The authors state the need for future research to devise improved techniques that can yield more satisfactory results~\cite{guglielmi2023using}. 

%\textbf{
Some research on game videos has looked specifically at identifying the occurrence of certain events in the videos. %}
Ringer et al. created a technique to detect highlights from video game streams by analyzing both game footage and footage of the streamer as they played~\cite{ringer_deep_2018}. They found that this combination of features was able to effectively identify highlights in game streams for the game \textit{Player
Unknown’s Battlegrounds}~\cite{ringer_deep_2018}.
 Roohi et al. created an annotation system to detect the emotions of game streamers over time through a neural network-based analysis of ``facial expressions, video transcript sentiment, voice emotion, and low-level audio features (pitch, loudness)'' and achieved accuracy scores that reached as high as 70.7\%~\cite{roohi_recognizing_2019}. 
Song et al. created a database of audiovisual data relating to player frustration when playing a game \cite{song_audiovisual_2019}. The authors extracted these features from recorded footage of participants playing a game and created a model to automatically detect moments of player frustration~\cite{song_audiovisual_2019}. This model achieved an unweighted average recall score of 60.3\% \cite{song_audiovisual_2019}.

These works are distinguished from ours in the sense that none of them are searching for occurrences of bugs. Additionally, these works all incorporated footage of the player on some level into their techniques, which is not required for our approach.

Cooper et al. devised a technique to detect duplicate bugs within video-based bug reports~\cite{cooper_it_2021}.  
Their technique was able to identify ``duplicate video-based bug reports within the top-2 candidate videos for 83\%'' of the evaluated tasks~\cite{cooper_it_2021}.
This work differs from ours in a couple of ways. First, the technique presented in this paper is not specific to games; it was evaluated on a set of six Android apps from a variety of categories such as productivity and finance~\cite{cooper_it_2021}. Second, the purpose of this technique is duplicate detection; rather than identifying whether a given video contains a bug, it instead returns a ranked list of other videos, where higher ranked videos ``are more likely to show the same bug'' as the input video~\cite{cooper_it_2021}.

%\textbf{
An additional body of research has explored bug identification and prevention in the domain of video games. %}
Truelove et al. expanded on the bug type taxonomy created by Lewis at al.~\cite{lewis_what_2010} and investigated the types of bugs that are fixed in game updates~\cite{truelove2021we}. They also surveyed game developers about the challenges they associated with the appearance of bugs~\cite{truelove2021we}. Survey respondents mentioned difficulties with reproducing bugs~\cite{truelove2021we}.
Looking at the issue of load testing video games, Tufano et al. implemented reinforcement learning to ``train an agent able to play the game as a human'' in order to identify bugs, particularly bugs related to drops in game performance~\cite{tufano_using_2022}. They found that their trained model was able to outperform a model that was only trained to play the game~\cite{tufano_using_2022}.

Our work is distinct from these works in the sense that neither work investigates gameplay videos. Truelove et al. focused on bugs described in game updates rather than on bug identification~\cite{truelove2021we}. While Tufano et al. looked at bug identification, this identification was done through the use of a reinforcement learning agent to simulate a human tester rather than through any video analysis~\cite{tufano_using_2022}.

\section{Methodology}
\label{sec:method}

\begin{figure*}[ht]
    \centering
    \includegraphics[width=0.80\linewidth]{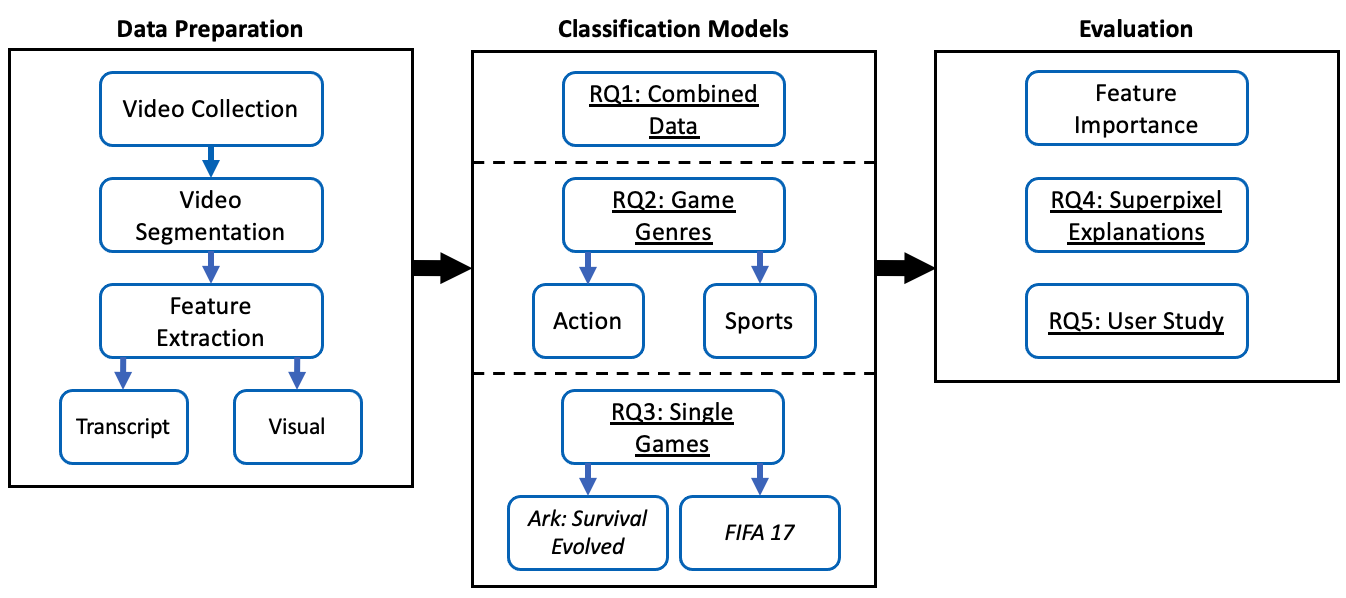}
    \caption{An outline of our methodology}
    \label{fig:method}
\end{figure*}

The following sections detail the research methodology of our study. Section 3.1 explains the video data we collected and the feature extraction techniques we applied. Section 3.2 describes the classification models we created. Section 3.3 looks at an analysis of the videos we examined, and Section 3.4 describes our user study.
Figure~\ref{fig:method} shows a summary outline of our methodology.

\subsection{Data Preparation}
\label{subsec:dataPrep}

\subsubsection{Video Collection}
\label{subsec:videoCollection}

Our dataset that we used to train and test our classification models was comprised of game play videos uploaded to YouTube. Prior work from Lin et al. had provided a dataset of online videos that showcased a bug~\cite{lin_identifying_2019}. We extracted the YouTube videos from this dataset for our study. Additionally, we also found a selection of bug videos from a cursory search of YouTube. This cursory search was performed through a simple keyword search, using terms that Lin et al. found to have a relation with bugs in games in their study~\cite{lin_identifying_2019}, including ``'bug'', ``glitch'', and ``hack''. Combining the videos from these two sources gave us an initial body of 891 game play videos. We then took a randomly selected sub-sample of these videos. Using a 95\% confidence interval with a 5\% margin of error, we selected 269 videos.

\begin{figure}[ht]
    \includegraphics[width=0.50\columnwidth]{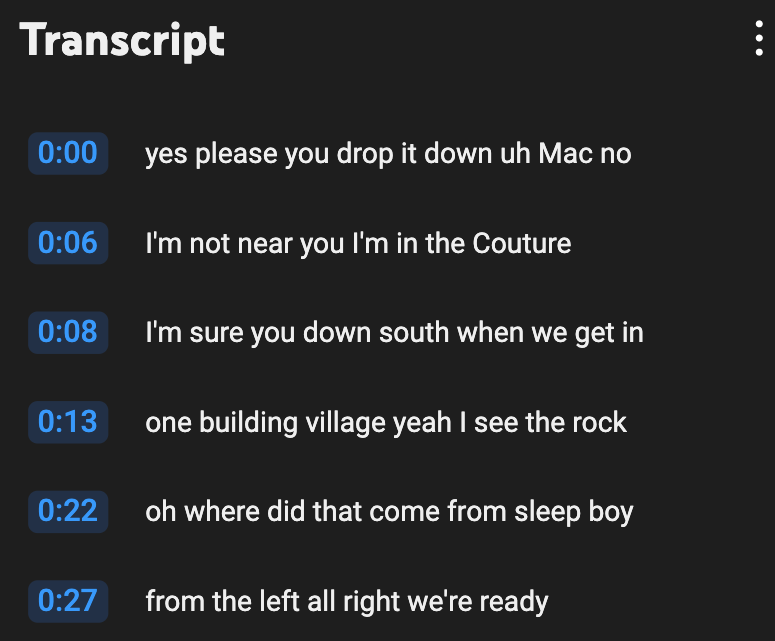}
    \caption{YouTube Transcript Example}
    \label{fig:captionExample}
\end{figure}

The next step was to divide the videos into segments. This division helped better identify the time in a video at which a bug appears, since using a model to classify a segment provides a more specific time frame than classifying the entire video. 
A segment also provides a possible start time and end time that would be found with a buggy segment as described by Lin. et al in their discussion of future work in bug identification in game videos~\cite{lin_identifying_2019}. 
When it came to how to divide the videos into segments, we looked to YouTube's video caption feature. YouTube allows users to add captions to their videos; the uploader of a video can either create the captions manually, or they can use YouTube's auto-caption feature to automatically generate captions from the video's speech~\cite{parton_video_2016}. If captions are enabled, users will see the captions appear on the screen as they view the video~\cite{parton_video_2016}. In addition to displaying the captions on the video, users also have access to a pane on the video page that lists a transcript of all captions in the video, which are paired alongside the timestamps at which the captions appear~\cite{hollett_how_2018M}. Figure \ref{fig:captionExample} shows an example of this caption transcript as it appears on the YouTube page for one of the videos from our dataset~\cite{pubg_spicyCat}. 
Users can utilize this pane to download the transcript information for videos~\cite{hollett_how_2018M}.

Because we wanted to use the transcripts of a video as a feature in our classifier, we decided to split our videos into smaller video segments based on the timestamps of the captions (see Section 3.1.2 for a deeper explanation of the transcript features, including the rationale for their inclusion). For the video from Figure \ref{fig:captionExample}, we split the video so that the first segment ran from the start of the video to the 0:06 mark, the second video segment began at the 0:06 mark and ran until the 00:13 timestamp, and so on. We decided to establish a minimum length of five seconds for each video segment in our dataset. In the event that a segment was less than 5 seconds, we combined it with the shortest adjacent segment. This was done iteratively to account for instances in which a newly combined segment was still less than 5 seconds long.

Since not all videos on YouTube make use of captions, however, we could not rely on YouTube as the sole source for video transcripts. For videos with no provided captions, we used Otter.ai, a service that automatically generates captions for videos ~\cite{noauthor_otterai_nodateM}. Past research evaluating Otter's caption generation have reported high levels of transcript accuracy for videos, reaching as high as 99.7\%~\cite{millett2021accuracy}. Uploading videos to Otter.ai produced transcripts documents similar to those from YouTube, each one consisting of a timestamp and the words spoken at that time. We used the timestamps from the Otter transcripts to split the videos in the same way done for the videos with YouTube captions.

One of the authors and two students then labelled the segments in our dataset based on whether the segment depicted the occurrence of a bug. Out of the 4,412 labelled segments in the dataset, there were fewer than 20 instances of disagreement.
During the course of our labelling, we discarded videos from our dataset that did not actually contain any gameplay footage, and videos that were no longer publicly available on YouTube.
In all, we ended up collected and labelling 4,412 video segments from 198 videos.

\subsubsection{Feature Extraction}

Our features came from two different information sources: the video transcripts and the visual data from the video images.

\textbf{Video Transcript Features:}
Prior research has incorporated video transcripts as a feature source for video classification tasks (e.g.,~\cite{chatbri2017automatic, liu2016multimodal, guglielmi2023using}).
To process the transcript text into our feature set, we followed the approach taken by Shah et al., who used the   
Universal Sentence Encoder~\cite{cer_universal_2018} to encode the transcript text of YouTube videos into 512-length numerical vectors~\cite{shah_automated_2019}. Though Shah et al. used these features to train a generative model rather than a classifier, we followed their approach because the the source of textual features they used were highly similar to ours; the authors also collected the transcripts of YouTube videos and used the provided timestamps to split the transcript text into phrases~\cite{shah_automated_2019}. We used the Universal Sentence Encoder to encode the transcript text of each video segment into a 512-length numerical vector. Each element in the vector was treated as an individual feature for our classifier.

\textbf{Visual Features:} Our process for extracting the visual features from the video segments was broadly modeled after the process utilized by Cooper et al., who used visual features for the task of duplicate detection in videos of bug reports, and were able to identify duplicates within the top-2 ranked candidate videos for 83\% of tasks~\cite{cooper_it_2021}. Our process follows the general steps they used and involves extracting and comparing frame images from our video dataset with one another. First, we extracted frame images from all the video segments at a rate of one frame per second. For each frame, we then used SimCLR~\cite{chen2020simple}, which was also used by Cooper et al.~\cite{cooper_it_2021}, to extract feature information from the image. More specifically, the output of the SimCLR feature extractor for every image is a vector of length 64.

While it would be possible to use these extracted vectors as features directly, these vectors represent individual frames, and our classifier was intended to work on segments of a video, i.e., on a sequence of multiple frames, meaning we needed a way to aggregate these frame vectors to represent the entire video segment.
Following the example of Cooper et al., we combined a Bag-of-Visual-Words approach with a TF-IDF approach~\cite{cooper_it_2021}. This approach has been shown to have work with high performance in prior papers~\cite{cooper_it_2021, kordopatis2019fivr}. The frame feature vectors were clustered, and the centroids of these clusters were then used to represent a different visual word~\cite{cooper_it_2021}. The frame image vectors in each video were then compared with the centroids, and through this we calculated a term frequency/document frequency measure for every centroid in the set.  

For our data, we used two different approaches to cluster our frame images. The first approach closely followed Cooper et al., where we simply used a \textit{k}-means algorithm to automatically generate our clusters and calculate the respective centroids. For our other approach, instead of using an automatic algorithm, we manually clustered our frames together so that all frames from the same video segment were grouped with one another. For the video segments that were labelled as depicting a bug, we manually examined the segment and extracted the frame in which the bug first manifests, and this frame was selected as that segment cluster's centroid. We decided to manually cluster frames in the same segment together to better help our models compare the more nuanced characteristics of buggy and non-buggy frames from the same video segment. This approach also helped ensure that the centroid image actually corresponded to the occurrence of a bug. A bug is by definition a break from the intended game experience, which meant buggy frames were frequently outliers compared to the other frames in that segment under the automatic approach. Manually identifying buggy centroids ensured that our models would investigate the characteristics of these buggy frames.     

Once our cluster sets were formed, we indexed the video segments in our data by comparing each frame to the centroids in our sets. For each frame in a video, the feature vector was matched to the closest centroid using cosine similarity~\cite{ye2011cosine}. To obtain the final visual features, a TF-IDF calculation occurs for every centroid in the BOVW. Where the term frequency (TF) is the frequency at which a centroid image is matched with a frame in the specific video segment, and the inverse document frequency corresponds with the number of times the centroid matched with a frame in the entire data set~\cite{cooper_it_2021}. This results in a vector for each video with a length equal to the number of clusters in the BOVW. When we manually grouped the frames by the segment, we ended up with 5,035 clusters. When using the automatic \textit{k}-means approach, we made sure to keep the same number of clusters for comparison. Each element of the vector was treated as an individual feature for the classification model.

\subsection{Classification Models}
\label{subsec:ClassModels}

For our classification model, we made use of a number of different model types, including: 
\begin{itemize} 
\item NeuralNetFastAI~\cite{howard2020fastai},
\item lightGBM~\cite{ke2017lightgbm},
\item XGBoost~\cite{chen2016xgboost},
\item Linear Model~\cite{skLearn_LinearModel},
\item Random Forest~\cite{skLearn_RandomForest}, and
\item A Weighted Ensemble that combined all our model types through ensemble selection~\cite{caruana2004ensemble}
\end{itemize}

\begin{figure}[ht]
    \includegraphics[width=0.80\columnwidth]{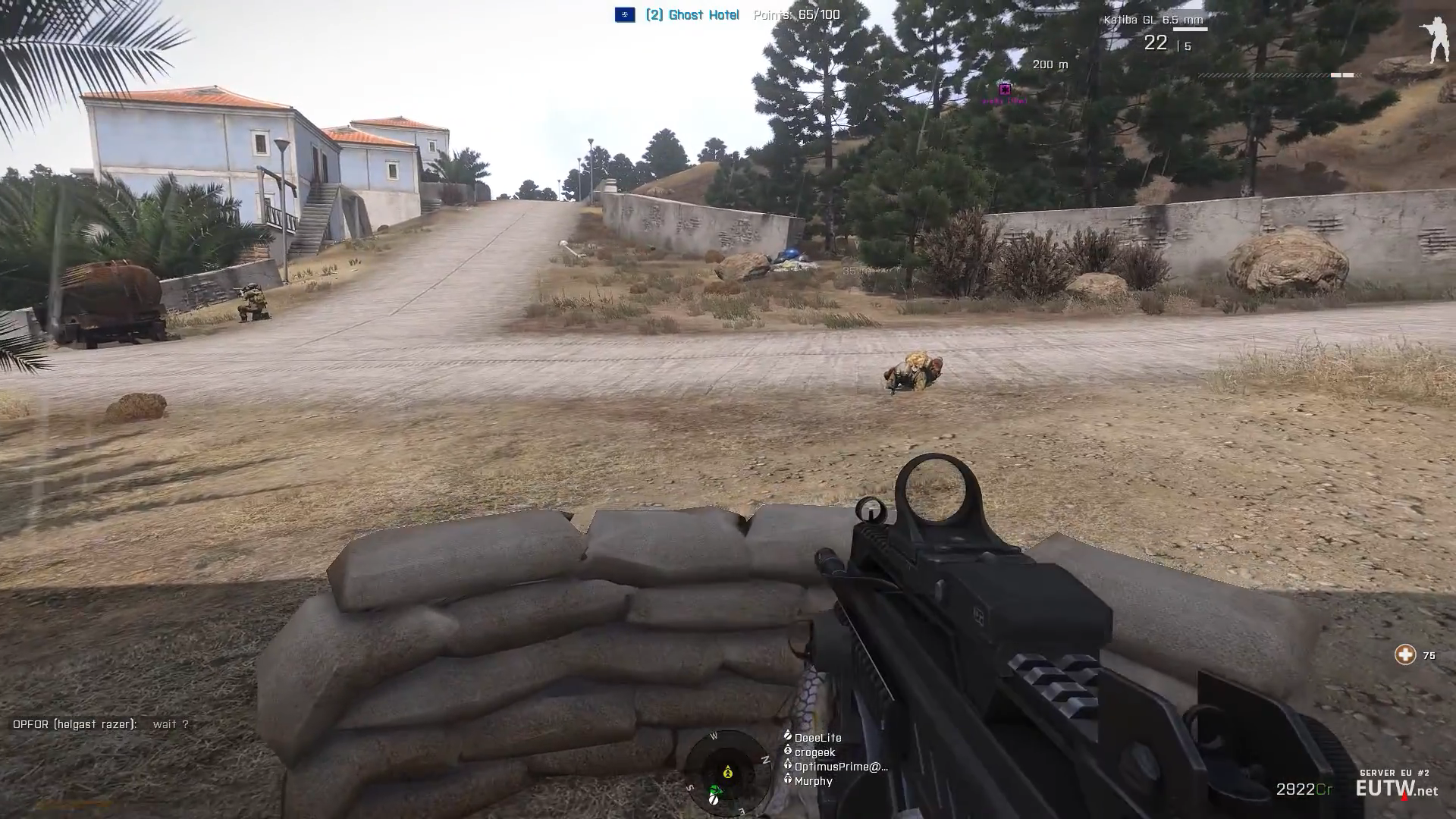}
    \caption{\textit{Arma 3} Screenshot} 
    \label{fig:arma3}
\end{figure}

\begin{figure}[ht]
    \includegraphics[width=0.80\columnwidth]{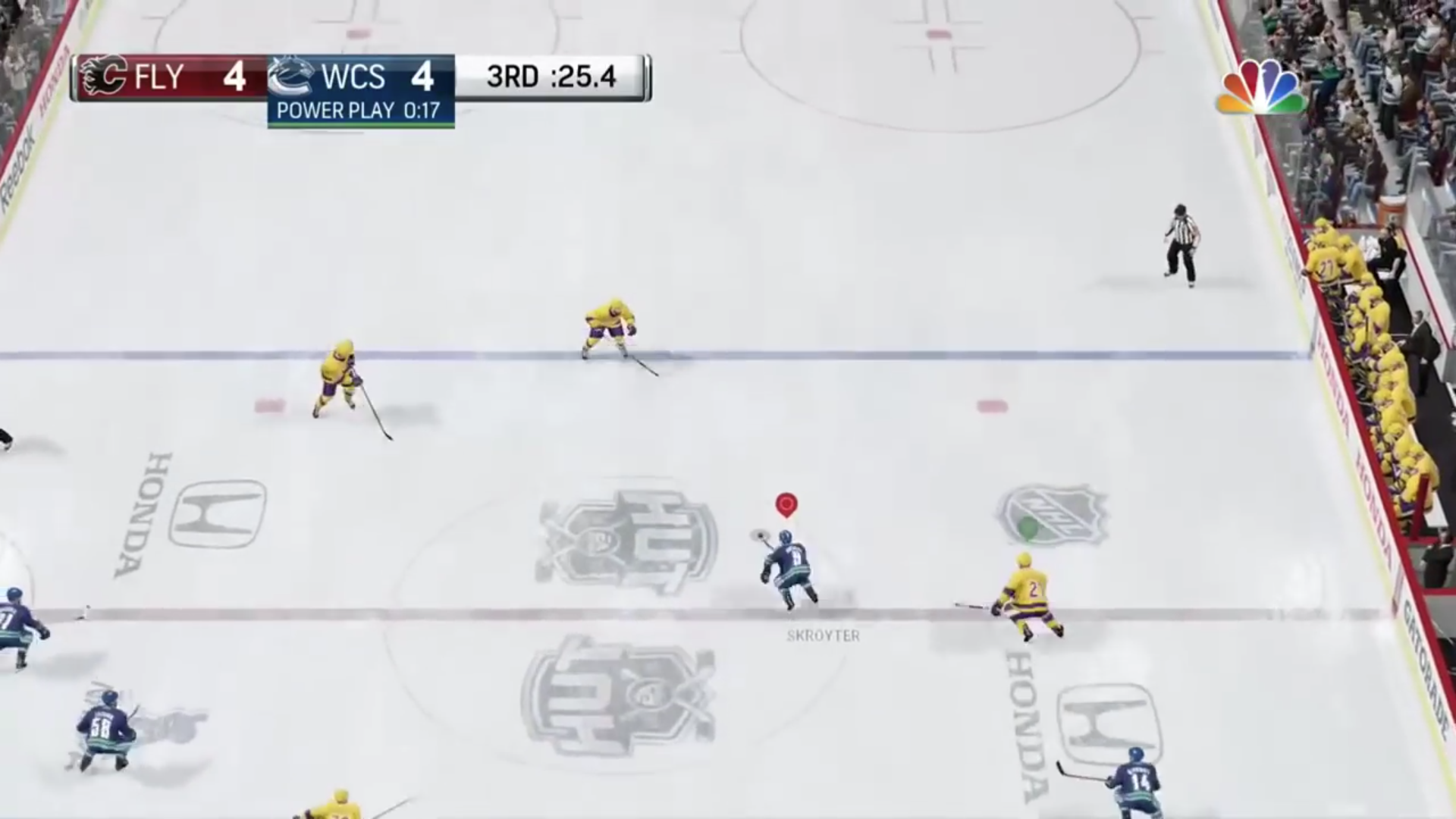}
    \caption{\textit{NHL 16} Screenshot} 
    \label{fig:nhl16}
\end{figure}

We performed parameter tuning as well for our models, applying different learning rates, numbers of epochs, etc. We ran our classifier on our full dataset of all video segments, but we also wanted to see how a classifier would perform on smaller, more specialized datasets.
The motivation for narrowing our focus in this respect was largely due to how diverse the game industry has become. The player view (any by extension, the visual and audio content) can vary significantly depending on the type of game being played. As an example, Figures \ref{fig:arma3} and \ref{fig:nhl16} shows screenshots from two different games captured by videos in our dataset (\cite{crogeekArma2M, eaNHL16M}). Figure \ref{fig:arma3} is from \textit{Arma 3}, and Figure \ref{fig:nhl16} is from \textit{NHL 16}. In a game like \textit{Arma 3}, the player is given a first-person perspective from behind the player character. The player's view is more-or-less constrained to the view of the player character. In a game like \textit{NHL 16}, however, the player view is high above the game world, allowing the player to see several characters at once.

The differences in the player's view between these games suggest that classification models using video features might perform better if they are trained and run on a subset of videos showing similar types of game play. In order to test this notion, we also trained and tested our models on different subsets of our dataset. The subsets are described in the following subsections.

\subsubsection{Game Genres}
\label{subsubsection: gameGenres}

Out of the 198 videos in our dataset, 96 were for games that simulated a real world sport. The remaining 102 videos were from games that heavily revolved around combat, such as shooting or attacking with handheld weapons. As such, we split the games depicted in these videos to two categories: sports games and action games. Of our 4,412 total video segments, 2,772 segments were from sports games and 1,589 were from action games. We created two data subsets for each of these game genres.

\subsubsection{Individual Games}
\label{subsubsection: indGames}

Since we were already splitting our segment dataset by game genre, we were curious about going one step further and examining if it was possible to train and test models on a single game. Doing so could be useful for developers who only work with a single game. A model specialized on a single game may be better able to recognize the unique characteristics of that game. Two evaluate this question, we took the two primary game genres we identified and found the game within each genre with the most segments in our data. The sports game with the most video segments was \textit{FIFA 17} with 952 segments. The action game with the most video segments was \textit{Ark: Survival Evolved} with 694 segments. We extracted these segments and created two data subsets for each of these two games.

\subsection{Explaining Predictions}
\label{subsec:explaining}

In addition to developing our classifiers, we also wanted to investigate the possible reasons why a classifier might decide to predict whether a video segment contained a bug. As such, we used LIME~\cite{ribeiro_why_2016} to identify superpixels--the specific regions of pixels in the image that supposedly held importance to the classifier when it made its prediction~\cite{stutz2018superpixels, ribeiro_why_2016}. To generate these superpixels, we used the Inception v3 image model~\cite{szegedy_going_2015} and fine-tuned it using the training data that was used to train our classifiers. We then tested the model on the buggy video segments from our test set and generated the superpixel maps. We examined the resulting maps to identify possible patterns in the superpixel regions the model identified. These patterns might provide insights into what characteristics of an image seem to have an association with bugs.

\subsection{Video Attributes}
\label{subsec: videoAttributes}
Going back to Figures~\ref{fig:arma3} and~\ref{fig:nhl16}, the idea that differences between particular games or genres might affect the performance of our models led us to consider whether any of those differences might be apparent in other parts of our data. In that sense, we also decided to examine the bug occurrence patterns in our dataset of videos to investigate whether there was any relation between the type of game and the patterns in which bugs appear in videos. To further explore the characteristics of different genres of video games, we defined following five attributes for a single video: 

\textbf{Total Number of Segments}: As explained in subsection~\ref{subsec:videoCollection}, we divided videos into several segments based on the transcript and timestamp feature. This attribute counts the total number of segments of a game video after division.

\textbf{Number of Buggy Segments}: Based on the prediction results, this attribute counts the total number of segments labeled as "showing bugs" for a single game video. 

\textbf{Buggy Segment Ratio}: To compare if one video is more likely to have a bug than other videos, we normalized each video's count of buggy segments by its total counts of buggy segments. This attribute calculates the number of buggy segments divided by total number of segments, returning the buggy segment ratio for each video. 

\textbf{Start Time Ratio}:  
We also wanted to study if videos from different games varied regarding the time at which the first bug appeared (i.e., do bugs tend to appear earlier in action game videos than in sports game videos?). Identifying potential patterns in when bugs might appear can provide an additional layer of information to developers to supplement the identification models. The start time Ratio attribute calculates the start time of the first buggy segment divided by the total duration of the video. The units of time were in seconds. 

\textbf{Number of Gaps}. It is possible that consecutive buggy segments were talking about the same bug. Therefore, it was also important to identify discrete buggy segments in a video. To achieve this, this attribute counts the number of gaps between two subsets of consecutive buggy segments. For example, if a video has a labeled sequence $0,1,1,0,0,0,1,0,0,1$ and $1$ indicates the buggy segments, the number of gaps will be $2$ in this case, as there are two consecutive sequences with $0$ bounded by $1$.

Based on the game genre categorization from subsection~\ref{subsubsection: gameGenres}, our goal was to check if there was any difference for each attribute stated above between two genres. In total, we have $102$ action games and $96$ sports games. First, for each attribute, we checked whether their values were normally distributed using a Kolmogorov Smirnov test~\cite{massey1951kolmogorov}. Since the p-values were all larger than $0.05$, we concluded that all values were normally distributed. Then, we conducted the parametric two sample t-test to check if there was any difference between action games and sports games for each attribute. For \emph{Start Time Ratio} and \emph{Number of Gaps}, we excluded videos with no buggy segments, which resulted in $69$ number of videos for both genres. Then, we followed the same procedure to check if the value of these two attributes are normally distributed, and then conduct the two sample t-test on each of them between action games and sports games. 

\subsection{User Study}
\label{sebsec:userStudy}

To evaluate the performance of our identification approach, we conducted a user study to examine the time and accuracy of humans manually searching for bugs. Participants were given a form containing five links to videos in our dataset. They were asked watch each video and to mark the start times and end times of any buggy moments they witnessed. All participants were graduate students. In total, we received 17 responses for each of the selected five links.

For each video, we collected a list of start time and end time buggy segments pairs from the participants. Then, we mapped these buggy segments to the actual division of segments from subsection~\ref{subsec:videoCollection}. To get the correct mapping, we checked if the actual buggy segment is the subset of one of the user inputted start time and end time pairs. If the user inputted time window is a subset of the actual buggy segment, we would also mark the time window of the actual buggy segment as buggy, so that the information would be consistent between user study and the mining result.

Next, we followed the same calculation procedure of attributes stated in subsection~\ref{subsec: videoAttributes}. Since not all attributes are meaningful in this comparison, and there is an redundancy between \emph{Number of Buggy Segments} and \emph{Buggy Segment Ratio} in the correlation aspect. In general, we selected two attributes for statistical analysis: \emph{Start Time Ratio},  and \emph{Buggy Segment Ratio}. To compare user study and mining result on the selected five videos, we combined the study result from the 17 participants by calculating the average for each selected attribute grouped by videos.

\section{Results}
\label{sec:results}
To evaluate the performance of our tool in RQ1, we took our dataset of 4,412 video segments we collected from Section~\ref{subsec:videoCollection} and randomly split the segments into a training set and a testing set. A portion of this training set was then excised and used as a validation set during the classifier training. 
When evaluating the tool's performance on segments from specific game genres (RQ2) and for individual games (RQ3), we used the subset datasets described in Sections~\ref{subsubsection: gameGenres} and~\ref{subsubsection: indGames}. Specifically, we created four subset datasets of our original dataset. For the genre analysis we created a dataset of \textit{Action} game segments and a dataset of \textit{Sports} game segments. For the individual game analysis, we created a dataset containing only video segments from the game \textit{Ark: Survival Evolved} and a dataset containing only video segments from the game \textit{FIFA 17}. We split each of these subset datasets into training and testing sets and ran them through our classifiers as well.

As a baseline, we implement the technique from Guglielmi et al. called GELID and run it on our main dataset as well as each of the subset datasets~\cite{guglielmi2023using}. Specifically, we run the segment classification step of GELID. Guglielmi et al. tested different configurations of features and classifiers on their test data~\cite{guglielmi2023using}; we use the configuration that achieved the highest performance in their paper and evaluate this configuration's performance on on our data.

In this section we report our results for the research questions.

\subsection{RQ1: How well does our tool perform in predicting bug occurrences in gameplay videos?}
\label{sec:result_RQ1}

Our goal was to identify bug occurrences from gameplay videos collected from YouTube. Table~\ref{tab:eval_full_all} shows the results of our classification models on our full set of test data.  
For the dataset using features from our manually designed segment clusters, the Linear Regression model performed the best, achieving an F1 score of 0.87. The automatically clustered dataset attained a marginally higher F1 score of 0.88, however. It achieved this score with the Weighted Ensemble model. 
These two classifiers both significantly outperform GELID, which only achieved an F1 score of 0.31 when run on our dataset. Our best-performing configuration outperformed GELID by 57 points. 

\observation{The Weighted Ensemble model from the automatically clustered segment configuration achieved the highest performance with an F1 score of 0.88, outperforming GELID by 57 points.}

% Please add the following required packages to your document preamble:
% \usepackage{multirow}
\begin{table}[t]
\begin{tabular}{lllll}
\hline
Technique                 & Best Model        & F1   & Precision & Recall \\ \hline
\multirow{9}{*}{Manual Clustering}    & NeuralNetFastAI   & 0.77 & 0.67      & 0.91   \\
                           & LightGBM          & 0.78 & 0.95      & 0.67   \\
                           & XGBoost           & 0.38 & 0.57      & 0.28   \\
                           & Linear Regression & 0.87 & 0.91      & 0.84   \\
                           & RandomForest      & 0.82 & 0.98      & 0.71   \\
                           & ExtraTrees        & 0.82 & 0.98      & 0.70   \\
                           & CatBoost          & 0.75 & 0.98      & 0.60   \\
                           & KNeighbors        & 0.64 & 0.62      & 0.67   \\
                           & Weighted Ensemble & 0.84 & 0.80      & 0.88   \\ \hline
\multirow{9}{*}{Automatic Clustering} & NeuralNetFastAI   & 0.78   & 0.67        & 0.92     \\
                           & LightGBM          & 0.77   & 0.98        & 0.64     \\
                           & XGBoost           & 0.36   & 0.82        & 0.23     \\
                           & Linear Regression & 0.86   & 0.93        & 0.80     \\
                           & RandomForest      & 0.81   & 1.00        & 0.68     \\
                           & ExtraTrees        & 0.81   & 1.00        & 0.67     \\
                           & CatBoost          & 0.71   & 1.00        & 0.55     \\
                           & KNeighbors        & 0.61   & 0.66        & 0.56     \\
                           & Weighted Ensemble & \textbf{0.88} & 0.91      & 0.85   \\ \hline
GELID                      &                   & 0.31 & 0.34      & 0.28 \\ \hline 
\end{tabular}
\caption{Performance of the different classification on the full dataset}
\label{tab:eval_full_all}
\end{table}

\subsection{RQ2: How does the tool perform when trained on footage from different genres of video games?}
\label{sec:result_RQ2}

Table~\ref{tab:eval_genre_all} presents the performance on the classifiers on the \textit{Action} games subset dataset and the \textit{Sports} game subset dataset. 
With the \textit{Action} models, the top performing model was the Weighted Ensemble model trained on the manually clustered dataset. This model achieved an F1 score of 0.84. For the \textit{Sports} models, the best performance was also from the Weighted Ensemble model, though this time it was the one trained on the automatically clustered dataset. This model earned an F1 score of 0.90.%}

For the \textit{Action} models, the top models from the automatic dataset were NeuralNetFastAI and the Weighted Ensemble, which both achieved an F1 score of 0.64, though this is score is a full 20 points lower than the top model from the manual set. There was a noteworthy difference in the top \textit{Sports} models as well; the Linear Regression model that achieved a top F1 score of 0.80, which is still 10 points lower than the top manual model. 
All clustering configurations for both genre subset datasets still outperform GELID, however, 
with the top \textit{Action} model outperforming GELID by 58 points and the top \textit{Sports} model outperforming it by 66 points.

The performance of the top models saw modest improvement when moving from the full dataset to the \textit{Sports} games subset dataset, going from 0.88 in Table~\ref{tab:eval_full_all} to 0.90 in Table~\ref{tab:eval_genre_all}. On the other hand, there was a drop in the top model performance when going from the full dataset to the \textit{Action} game subset dataset, where the top F1 score was only 0.84. 
Using a model trained specifically on video segments from within a single game genre did not appear to lead to a universal performance boost when tested on other video segments from the same game genre. 

\observation{For the \textit{Action} game subset dataset, the top-performing model achieved an F1 score of 0.84. For the \textit{Sports} games, the top model achieved an F1 score of 0.90, outperforming GELID by 58 points and 66 points, respectively.}

\textbf{Video Attributes}: We also wanted to check if there was any difference between the video attributes between action and sports games (see Section~\ref{subsec: videoAttributes} for details). We found that there were statistically significant differences on the buggy ratio and start time ratio between two genres with a small cohen's D effect size (0.36). Action game videos tended to have a greater buggy ratio and earlier start times compared to sports game videos. The means of the action games buggy ratio and start time ratio were $0.43$, and $0.18$, respectively, while sports games had a mean of $0.29$, and $0.27$, respectively.

% Please add the following required packages to your document preamble:
% \usepackage{multirow}
\begin{table}[]
\begin{tabular}{llllll}
\hline
Game Genre                        & Technique                             & Best Model        & F1            & Precision & Recall \\ \hline
\multirow{19}{*}{\textit{Action}} & \multirow{9}{*}{Manual Clustering}    & NeuralNetFastAI   & 0.83          & 0.84      & 0.82   \\
                                  &                                       & LightGBM          & 0.75          & 0.98      & 0.60   \\
                                  &                                       & XGBoost           & 0.60          & 0.71      & 0.52   \\
                                  &                                       & Linear Regression & 0.82          & 0.92      & 0.74   \\
                                  &                                       & RandomForest      & 0.76          & 0.95      & 0.63   \\
                                  &                                       & ExtraTrees        & 0.76          & 0.95      & 0.63   \\
                                  &                                       & CatBoost          & 0.70          & 1.00      & 0.54   \\
                                  &                                       & KNeighbors        & 0.64          & 0.49      & 0.91   \\
                                  &                                       & Weighted Ensemble & \textbf{0.84} & 0.87      & 0.82   \\ \cline{2-6} 
                                  & \multirow{9}{*}{Automatic Clustering} & NeuralNetFastAI   & 0.64          & 0.71      & 0.59   \\
                                  &                                       & LightGBM          & 0.13          & 0.70      & 0.07   \\
                                  &                                       & XGBoost           & 0.07          & 0.25      & 0.04   \\
                                  &                                       & Linear Regression & 0.48          & 0.79      & 0.35   \\
                                  &                                       & RandomForest      & 0.04          & 1.00      & 0.02   \\
                                  &                                       & ExtraTrees        & 0.00          & 0.00      & 0.00   \\
                                  &                                       & CatBoost          & 0.02          & 1.00      & 0.01   \\
                                  &                                       & KNeighbors        & 0.25          & 0.93      & 0.15   \\
                                  &                                       & Weighted Ensemble & 0.64          & 0.71      & 0.59   \\ \cline{2-6} 
                                  & \multicolumn{2}{l}{GELID}                                 & 0.26          & 0.33      & 0.21   \\ \hline
\multirow{19}{*}{\textit{Sports}} & \multirow{9}{*}{Manual Clustering}    & NeuralNetFastAI   & 0.70          & 0.60      & 0.84   \\
                                  &                                       & LightGBM          & 0.72          & 0.93      & 0.59   \\
                                  &                                       & XGBoost           & 0.33          & 0.70      & 0.22   \\
                                  &                                       & Linear Regression & 0.80          & 0.95      & 0.69   \\
                                  &                                       & RandomForest      & 0.75          & 1.00      & 0.60   \\
                                  &                                       & ExtraTrees        & 0.75          & 1.00      & 0.60   \\
                                  &                                       & CatBoost          & 0.60          & 0.97      & 0.44   \\
                                  &                                       & KNeighbors        & 0.24          & 1.00      & 0.14   \\
                                  &                                       & Weighted Ensemble & 0.79          & 0.82      & 0.77   \\ \cline{2-6} 
                                  & \multirow{9}{*}{Automatic Clustering} & NeuralNetFastAI   & 0.84          & 0.80      & 0.90   \\
                                  &                                       & LightGBM          & 0.26          & 0.93      & 0.15   \\
                                  &                                       & XGBoost           & 0.28          & 0.79      & 0.17   \\
                                  &                                       & Linear Regression & 0.84          & 0.89      & 0.80   \\
                                  &                                       & RandomForest      & 0.80          & 1.00      & 0.67   \\
                                  &                                       & ExtraTrees        & 0.80          & 1.00      & 0.67   \\
                                  &                                       & CatBoost          & 0.82          & 1.00      & 0.69   \\
                                  &                                       & KNeighbors        & 0.46          & 0.90      & 0.31   \\
                                  &                                       & Weighted Ensemble & \textbf{0.90} & 0.92      & 0.89   \\ \cline{2-6} 
                                  & \multicolumn{2}{l}{GELID}                                 & 0.24          & 0.24      & 0.24   \\ \hline
\end{tabular}
\caption{Performance of classification techniques on the \textit{Action} and \textit{Sports} games subsets}
\label{tab:eval_genre_all}
\end{table}

\subsection{RQ3: How does the tool perform when trained on footage from a single game?}
\label{sec:result_RQ3}

Table~\ref{tab:eval_game_all} presents the performance of the classifiers on the \textit{Ark: Survival Evolved} subset dataset and the \textit{FIFA 17} subset dataset. 
With \textit{Ark: Survival Evolved}, the highest overall performing model was the Weighted Ensemble model that was trained on the automatically clustered dataset, achieving an F1 score of 0.73. For \textit{FIFA 17}, the Weighted Ensemble model from the manually clustered dataset attained a top F1 score of 0.91. %}
For the \textit{Ark: Survival Evolved} subset dataset, the top-performing model outperformed GELID by 44 points. 
Notably, for the \textit{FIFA 17} subset dataset, GELID attained an F1 score of 0.62; our top model outperformed GELID by 29 points. This is smaller gain compared to the other datasets, but GELID's performance on this data was still well below what was achieved by our top-performing \textit{FIFA 17} model. 

The F1 score for the \textit{FIFA 17} model saw a very marginal increase over the top F1 score achieved for sports games in general, going from 0.90 in Table~\ref{tab:eval_genre_all} to 0.91 in Table~\ref{tab:eval_game_all}.
On the other hand, training and testing specifically on \textit{Ark: Survival Evolved} did not end up leading to improved performance over training and testing on action games more generally; instead the top F1 score in Table~\ref{tab:eval_genre_all} dropped from 0.84 to the 0.73 seen in Table~\ref{tab:eval_game_all}. At this point there does not appear to be a clear upward or downward trend in performance when continuing to move from a more general domain to a more specific one.

\observation{For the \textit{Ark: Survival Evolved} subset dataset, the top-performing model achieved an F1 score of 0.73. For the \textit{FIFA 17} subset dataset, the top model achieved an F1 score of 0.91, outperforming GELID by 44 points and 29 points, respectively.}

% Please add the following required packages to your document preamble:
% \usepackage{multirow}
\begin{table}[]
\begin{tabular}{llllll}
\hline
Game Title                                       & Technique                             & Classification Model & F1            & Precision & Recall \\ \hline
\multirow{19}{*}{\textit{Ark: Survival Evolved}} & \multirow{9}{*}{Manual Clustering}    & NeuralNetFastAI      & 0.64          & 0.60      & 0.69   \\
                                                 &                                       & LightGBM             & 0.13          & 0.50      & 0.08   \\
                                                 &                                       & XGBoost              & 0.16          & 0.17      & 0.15   \\
                                                 &                                       & Linear Regression    & 0.43          & 0.73      & 0.31   \\
                                                 &                                       & RandomForest         & 0.00          & 0.00      & 0.00   \\
                                                 &                                       & ExtraTrees           & 0.00          & 0.00      & 0.00   \\
                                                 &                                       & CatBoost             & 0.00          & 0.00      & 0.00   \\
                                                 &                                       & KNeighbors           & 0.33          & 0.20      & 1.00   \\
                                                 &                                       & Weighted Ensemble    & 0.64          & 0.60      & 0.69   \\ \cline{2-6} 
                                                 & \multirow{9}{*}{Automatic Clustering} & NeuralNetFastAI      & 0.64          & 0.60      & 0.69   \\
                                                 &                                       & LightGBM             & 0.00          & 0.00      & 0.00   \\
                                                 &                                       & XGBoost              & 0.31          & 0.46      & 0.23   \\
                                                 &                                       & Linear Regression    & 0.41          & 0.88      & 0.27   \\
                                                 &                                       & RandomForest         & 0.00          & 0.00      & 0.00   \\
                                                 &                                       & ExtraTrees           & 0.00          & 0.00      & 0.00   \\
                                                 &                                       & CatBoost             & 0.00          & 0.00      & 0.00   \\
                                                 &                                       & KNeighbors           & 0.44          & 0.29      & 0.92   \\
                                                 &                                       & Weighted Ensemble    & \textbf{0.73} & 0.69      & 0.77   \\ \cline{2-6} 
                                                 & \multicolumn{2}{l}{GELID}                                    & 0.29          & 0.27      & 0.31   \\ \hline
\multirow{19}{*}{\textit{FIFA 17}}               & \multirow{9}{*}{Manual Clustering}    & NeuralNetFastAI      & 0.88          & 0.82      & 0.94   \\
                                                 &                                       & LightGBM             & 0.79          & 0.89      & 0.71   \\
                                                 &                                       & XGBoost              & 0.64          & 0.82      & 0.53   \\
                                                 &                                       & Linear Regression    & 0.90          & 1.00      & 0.82   \\
                                                 &                                       & RandomForest         & 0.81          & 0.96      & 0.71   \\
                                                 &                                       & ExtraTrees           & 0.81          & 0.96      & 0.71   \\
                                                 &                                       & CatBoost             & 0.80          & 0.92      & 0.71   \\
                                                 &                                       & KNeighbors           & 0.74          & 0.74      & 0.74   \\
                                                 &                                       & Weighted Ensemble    & \textbf{0.91} & 0.94      & 0.88   \\ \cline{2-6} 
                                                 & \multirow{9}{*}{Automatic Clustering} & NeuralNetFastAI      & 0.85          & 0.79      & 0.91   \\
                                                 &                                       & LightGBM             & 0.79          & 0.89      & 0.71   \\
                                                 &                                       & XGBoost              & 0.62          & 0.81      & 0.50   \\
                                                 &                                       & Linear Regression    & 0.88          & 0.91      & 0.85   \\
                                                 &                                       & RandomForest         & 0.77          & 0.96      & 0.65   \\
                                                 &                                       & ExtraTrees           & 0.77          & 0.96      & 0.65   \\
                                                 &                                       & CatBoost             & 0.80          & 0.92      & 0.71   \\
                                                 &                                       & KNeighbors           & 0.43          & 0.83      & 0.29   \\
                                                 &                                       & Weighted Ensemble    & 0.88          & 0.88      & 0.88   \\ \cline{2-6} 
                                                 & \multicolumn{2}{l}{GELID}                                    & 0.62          & 0.55      & 0.71   \\ \hline
\end{tabular}
\caption{Performance of classification techniques on the Ark: Survival Evolved and FIFA 17 subsets}
\label{tab:eval_game_all}
\end{table}

\subsection{RQ4: Is it possible to identify which characteristics of visual
video frames have an association with the prediction of a bug?}

We performed a feature importance analysis on our prediction models and found that the visual frame features consistently held a significantly greater amount of importance over the transcript features across our different models and training sets. However, when it comes to explaining our models' predictions, there is not much insight to be gleaned from the fact that a particular cluster in our Bag-of-Visual-Words set had a slightly higher TF-IDF than another. This is why we look to the generated pixel maps for this part of our analysis.

\begin{figure}[h]
    %\begin{subfigure}{0.5\textwidth}
    \includegraphics[width=.48\textwidth]{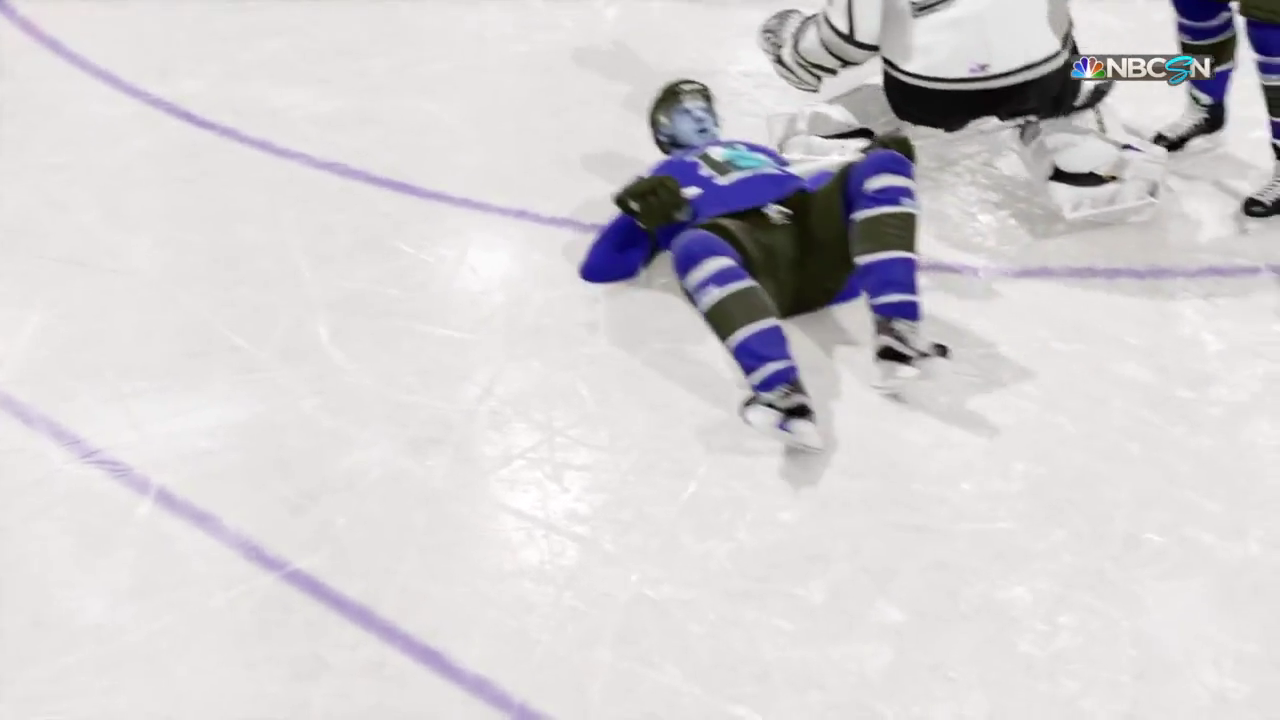}
    %\caption{Original Game Image}\hfill
    %\end{subfigure}
    %\begin{subfigure}{0.5\textwidth}
    \includegraphics[width=.48\textwidth]{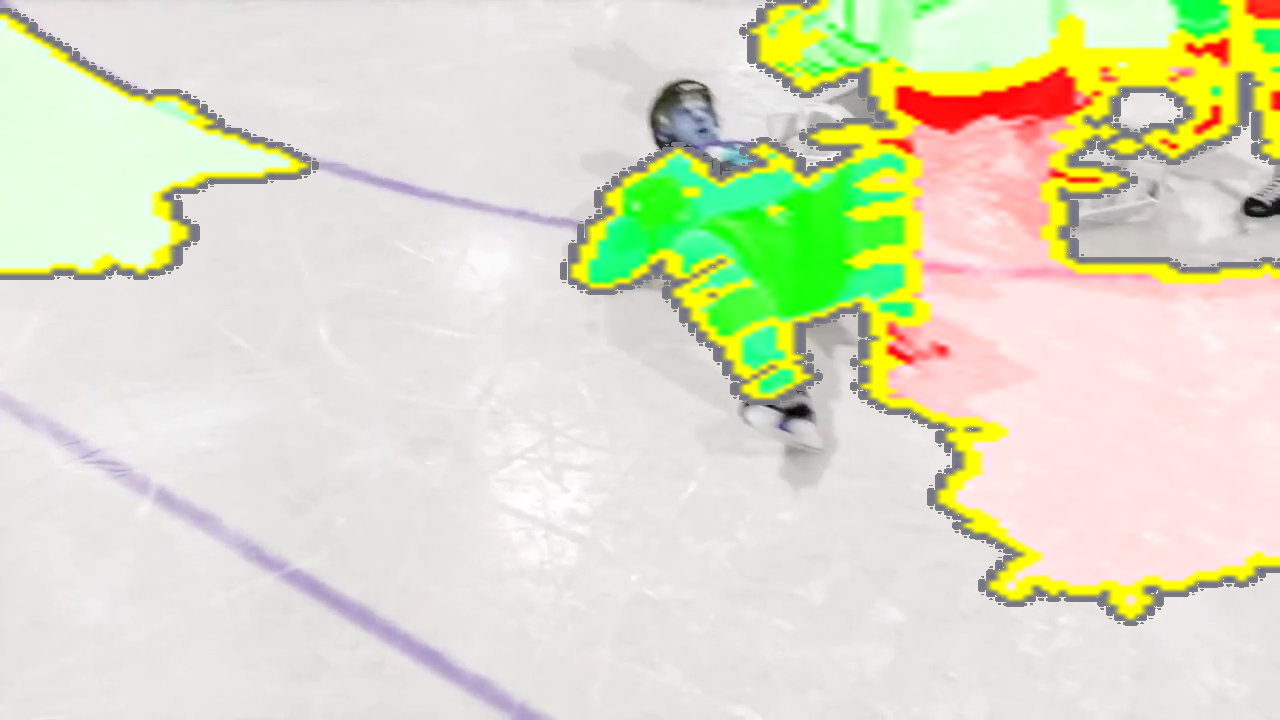}\hfill
    %\caption{pixel mapping}
    %\end{subfigure}
    \caption{Original Game Image (left) and the pixel mapping (right)}\label{fig:exampleMap}
\end{figure}

Figure~\ref{fig:exampleMap} shows an example of one of the pixel maps we got from our image model. The top image was a centroid image from one of our video segments. It was manually selected as a centroid after we verified that the frame did indeed depict a bug--the hockey player is stiff on the ice and unnaturally sliding down the rink. The resulting mapping indicates which parts of the image the model believed to have an effect on the ultimate decision to predict whether it contained a bug or not. The regions in green are areas that apparently had a positive influence on the model's classification, while the areas in red are the sports that worked against the ultimate decision. For this paper, we are not specifically concerned over whether an area is red or green as much as we are interested in what areas end up being covered at all. In this instance, the majority of the sliding player is covered in green, indicating that something about the appearance of this character led to the prediction's classification.

We examined the buggy segment centroids in our test data, categorizing them into three groups: frame images where the bug can be identified by looking at the frame on its own; frame images where the bug becomes apparent when comparing it to an adjacent frame, and frame images where the image does not relay any information about the type of bug being experienced. We put these frame images through the model, and examined them to try and identify any noteworthy patterns. Such patterns might provide insight into whether certain characteristics of images may have an association with the presence if bugs. In this section, we present a few simple patterns that we noticed during our examination.

\begin{figure}[h]
    \begin{subfigure}{1\textwidth}
    \includegraphics[width=.48\textwidth]{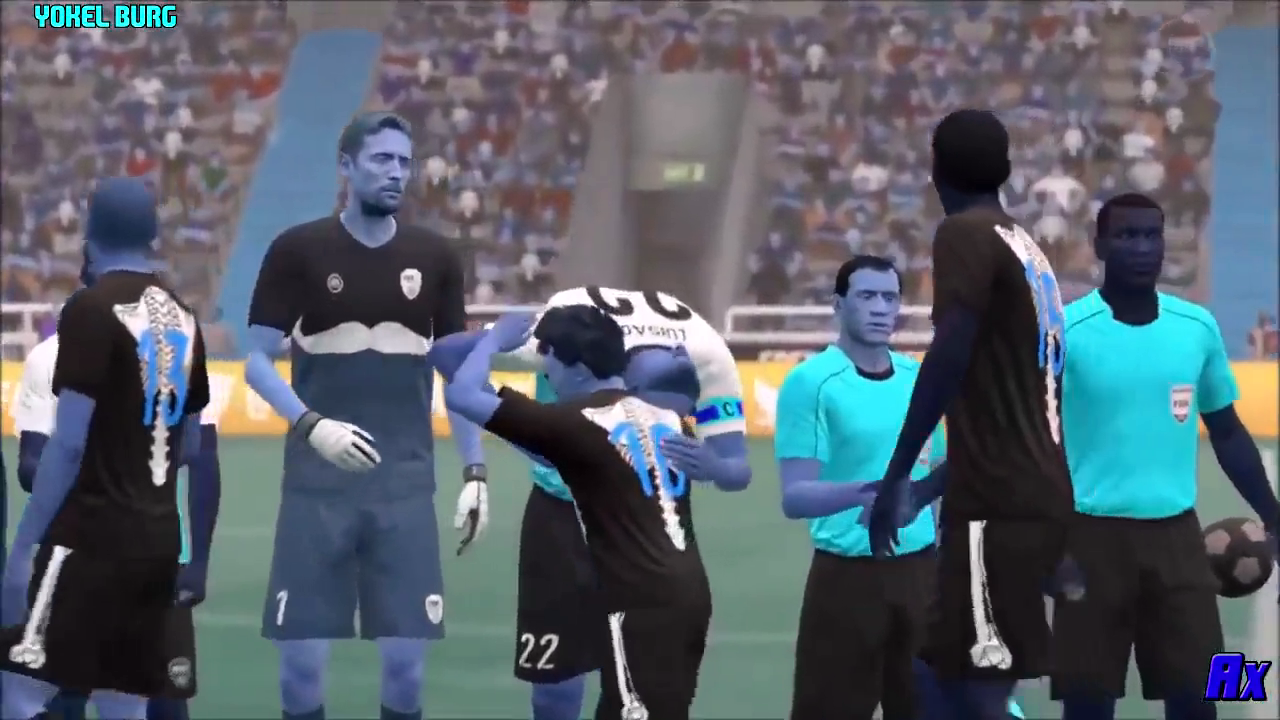}\hfill
    \includegraphics[width=.48\textwidth]{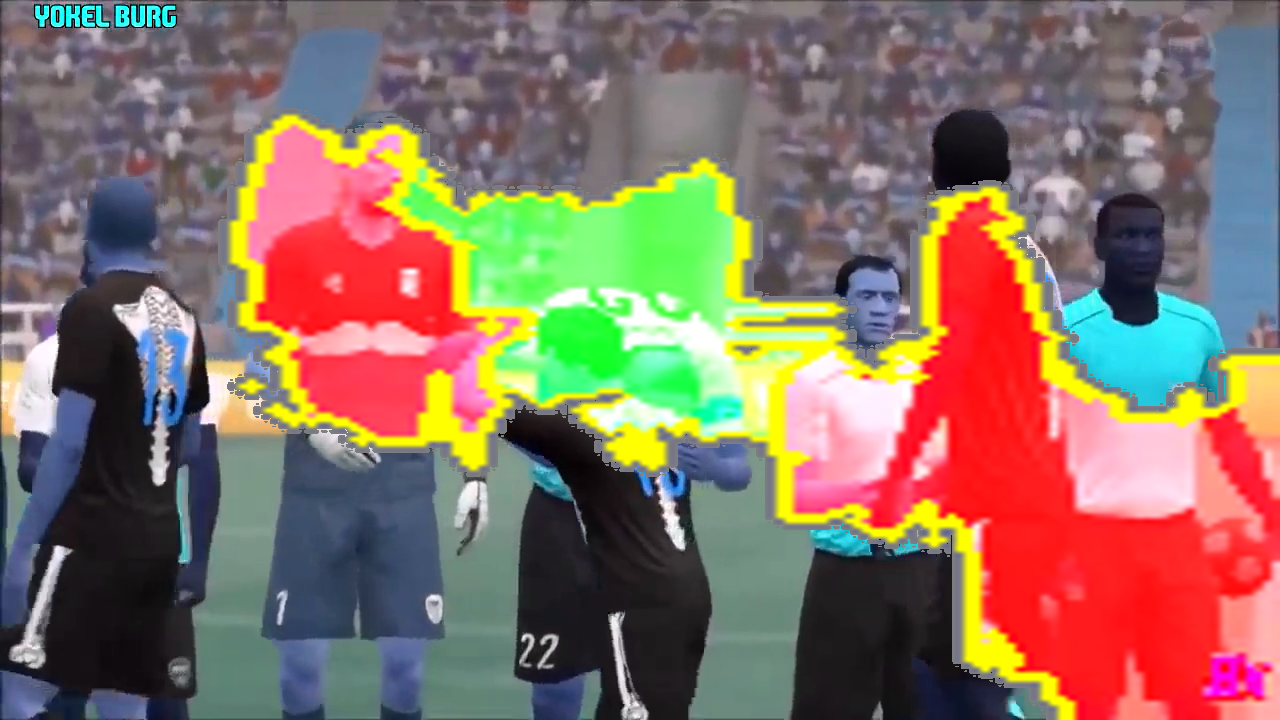}\hfill
    \caption{Soccer players gathered closely}\hfill
    \label{fig:bunchaSoccer}\hfill
    \end{subfigure}
    \begin{subfigure}{1\textwidth}
    \includegraphics[width=.48\textwidth]{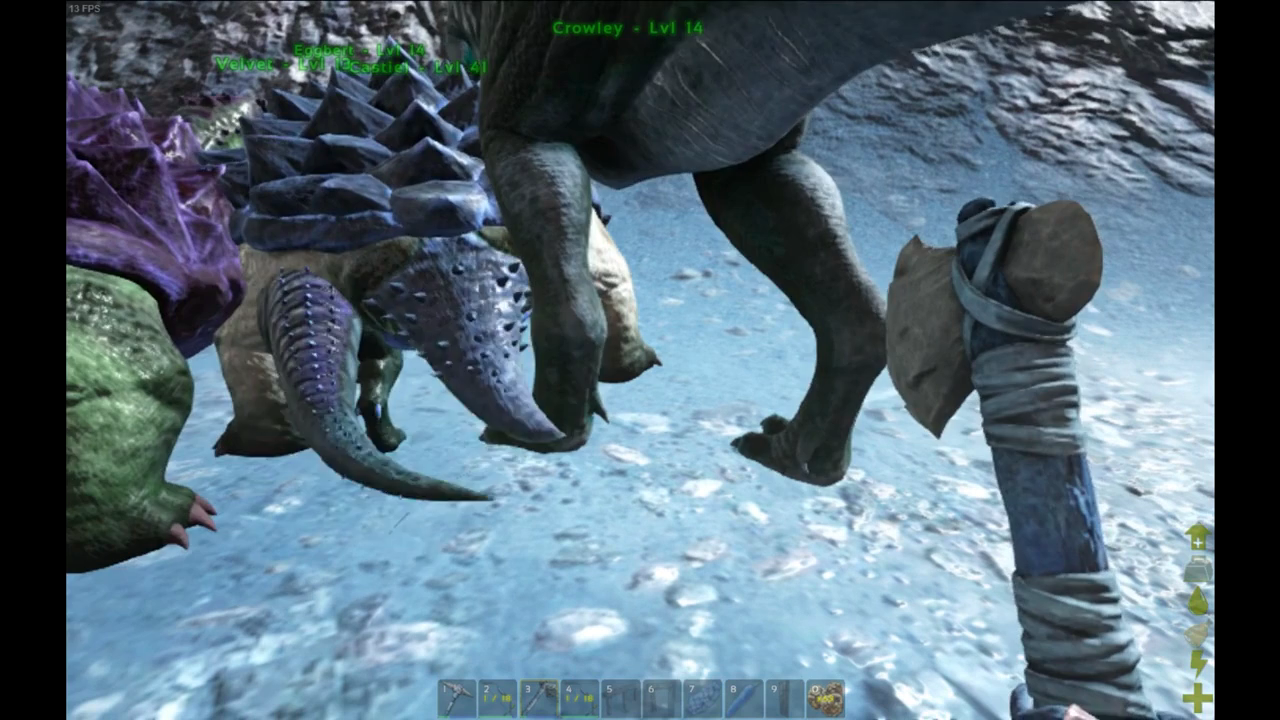}\hfill
    \includegraphics[width=.48\textwidth]{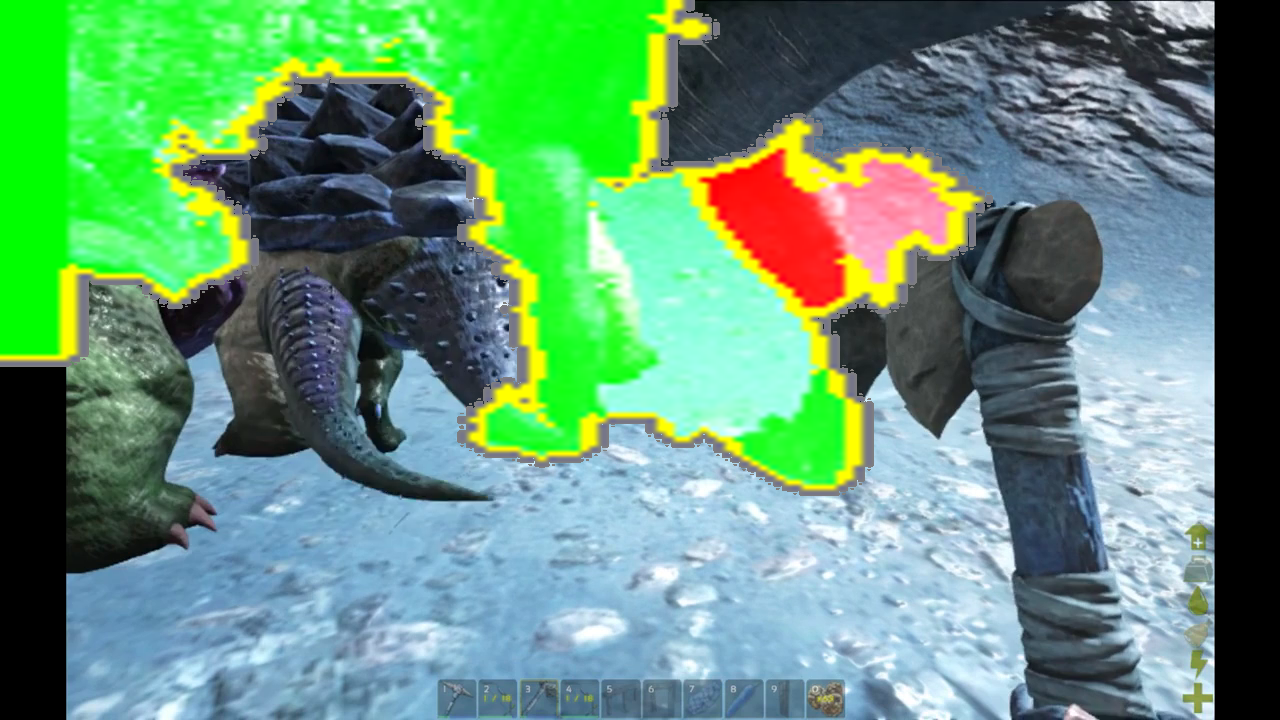}\hfill
    \caption{Bunched up dinosaurs}\hfill
    \label{fig:bunchaDinos}
    \end{subfigure}
    \vspace*{-3mm}
    \caption{Mappings of closely gathered characters}
    \label{fig:bunchedMap}
\end{figure}

\textbf{Closely Gathered Characters:} We found that the model had a tendency to highlight sections where multiple characters are positioned close together, such as in the example from Figure~\ref{fig:bunchaSoccer}, where two soccer players are unnaturally squished up together--the two players are highlighted green where they make contact with each other. While most examples involved human characters, it was possible for other types of characters to be highlighted as well, such as the tightly packed dinosaurs in Figure~\ref{fig:bunchaDinos}. Overall, the model seemed to have a sensitivity to people and to a lesser extent, animal characters.

\begin{figure}[h]
    \begin{subfigure}{1\textwidth}
    \includegraphics[width=.48\textwidth]{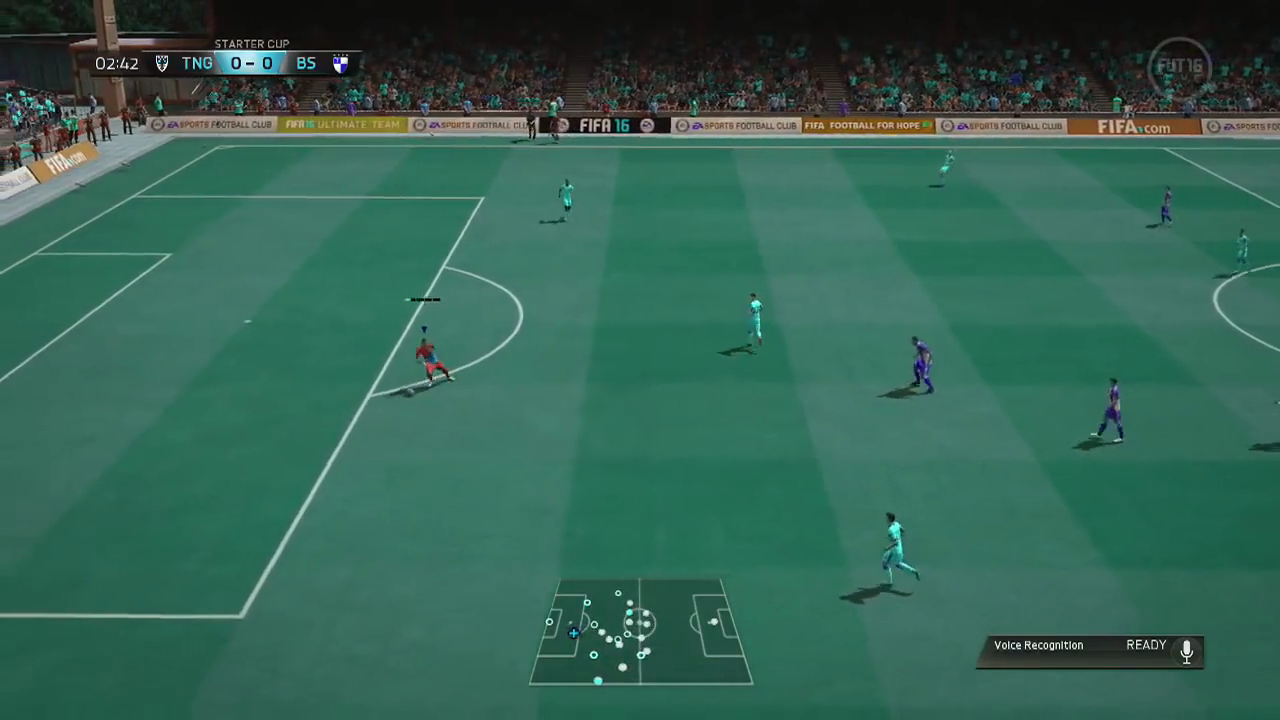}\hfill
    \includegraphics[width=.48\textwidth]{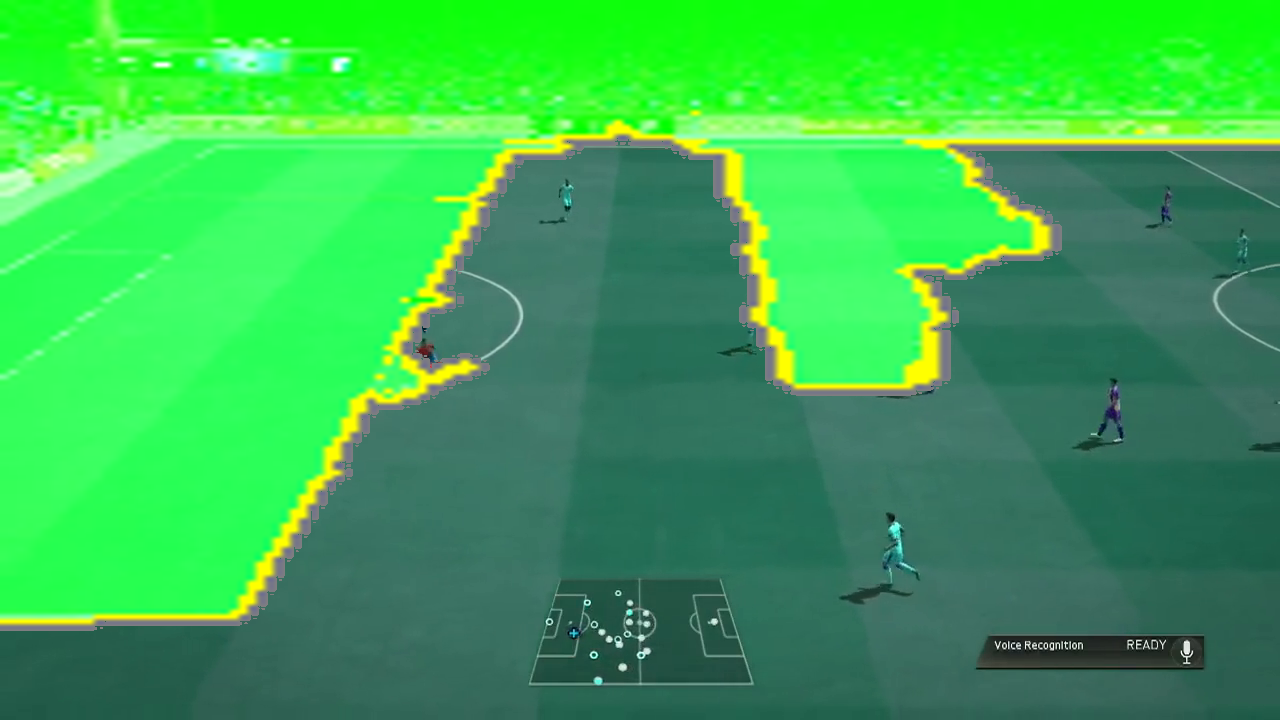}\hfill
    \caption{Mapping that follows lines on the field}\hfill
    \label{fig:shapeSoccer}\hfill
    \end{subfigure}
    \begin{subfigure}{1\textwidth}
    \includegraphics[width=.48\textwidth]{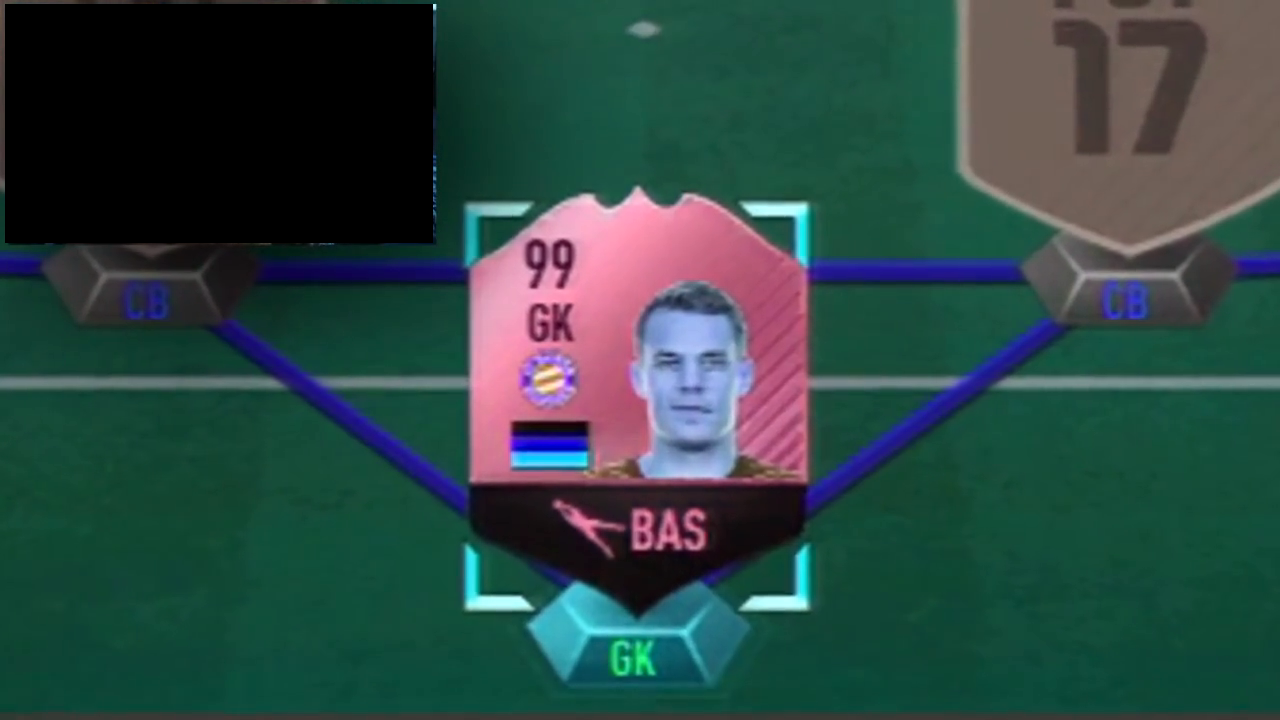}\hfill
    \includegraphics[width=.48\textwidth]{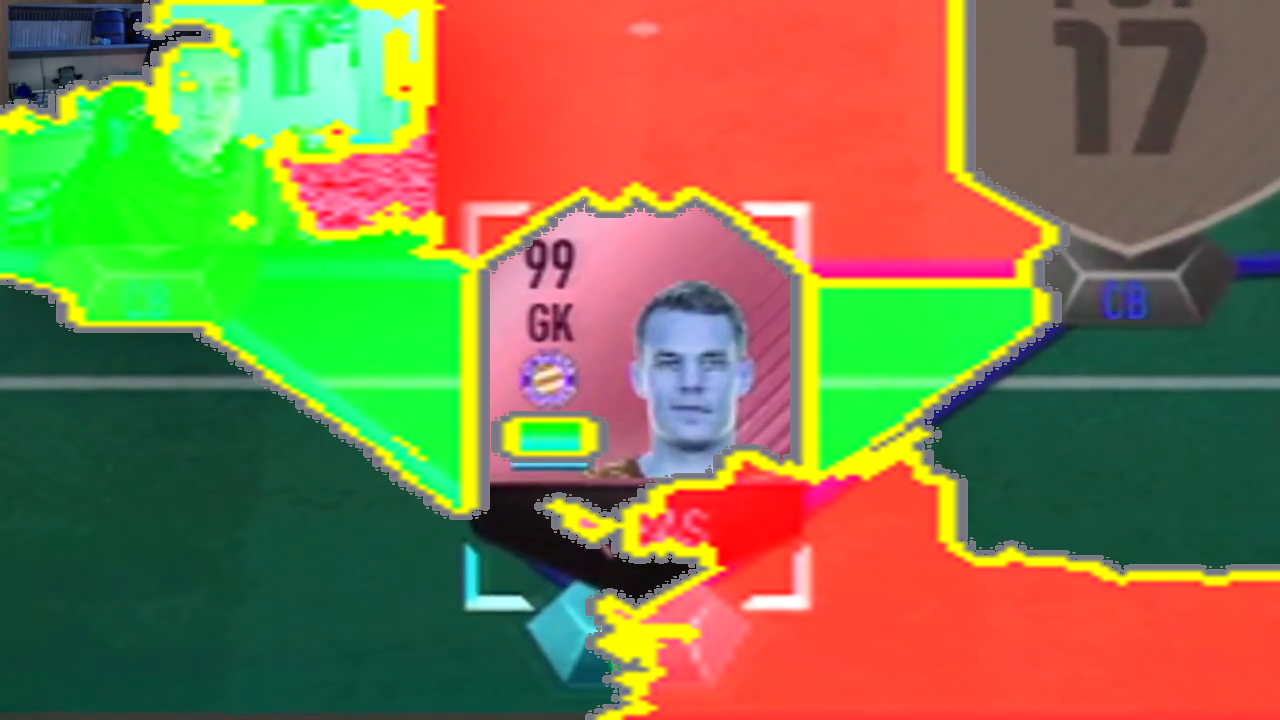}\hfill
    \caption{Mapping that follows shape of UI box}\hfill
    \label{fig:shapeUI}
    \end{subfigure}
    \vspace*{-10mm}
    \caption{Mappings showing an inclination to follow lines and shapes}
    \label{fig:shapesMap}
\end{figure}

\textbf{Bold Shapes \& Straight Lines:} This is an instance of the model perhaps not currently being optimized to handle certain types of images. More specifically, it seemed to get distracted by bold shapes and prominent straight lines in the absence of other more intricate objects in the frame. This was more common in sports games, since there were multiple instances in which the pixel mapping ended up following field and boundary lines, as seen in Figure~\ref{fig:shapeSoccer}. The model also had this problem in frame images featuring prominent UI elements and menus; as seen in Figure~\ref{fig:shapeUI}, the mapping seems to be more interested in fitting around the shape of the UI box than with the information depicted within it (which in this case is the actual source of the bug).

\begin{figure}[h]
    \begin{subfigure}{1\textwidth}
    \includegraphics[width=.48\textwidth]{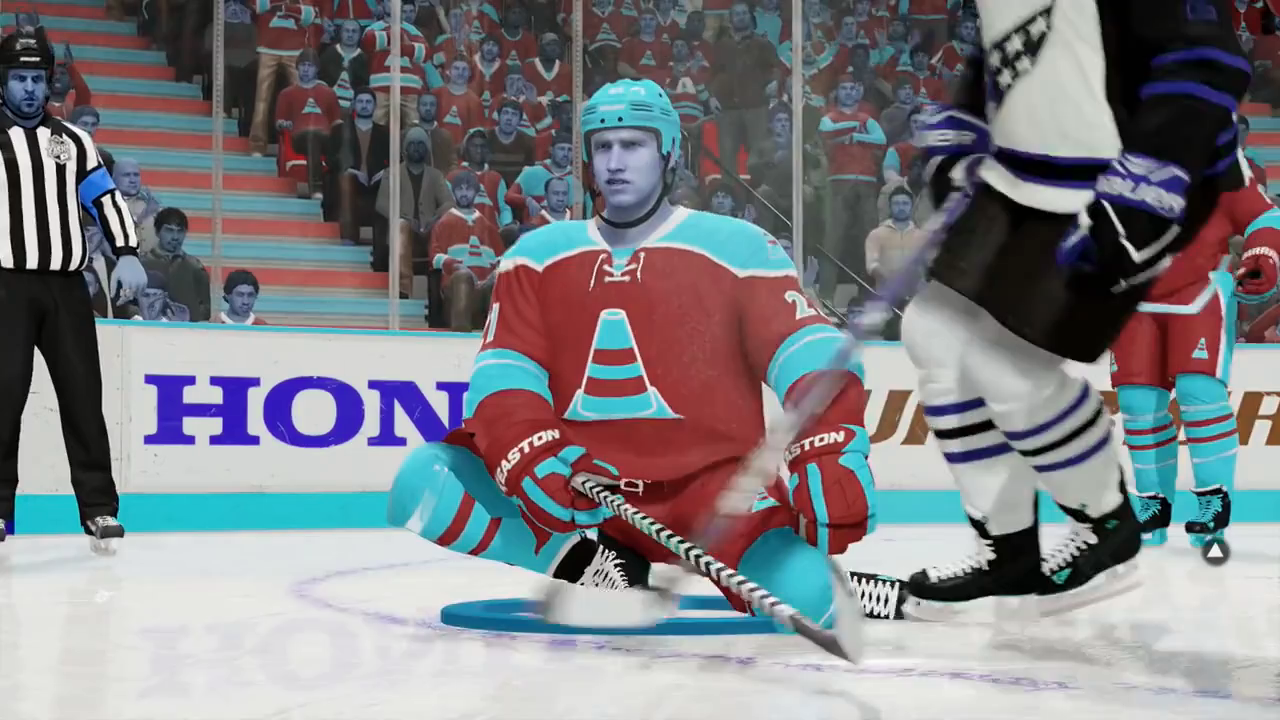}\hfill
    \includegraphics[width=.48\textwidth]{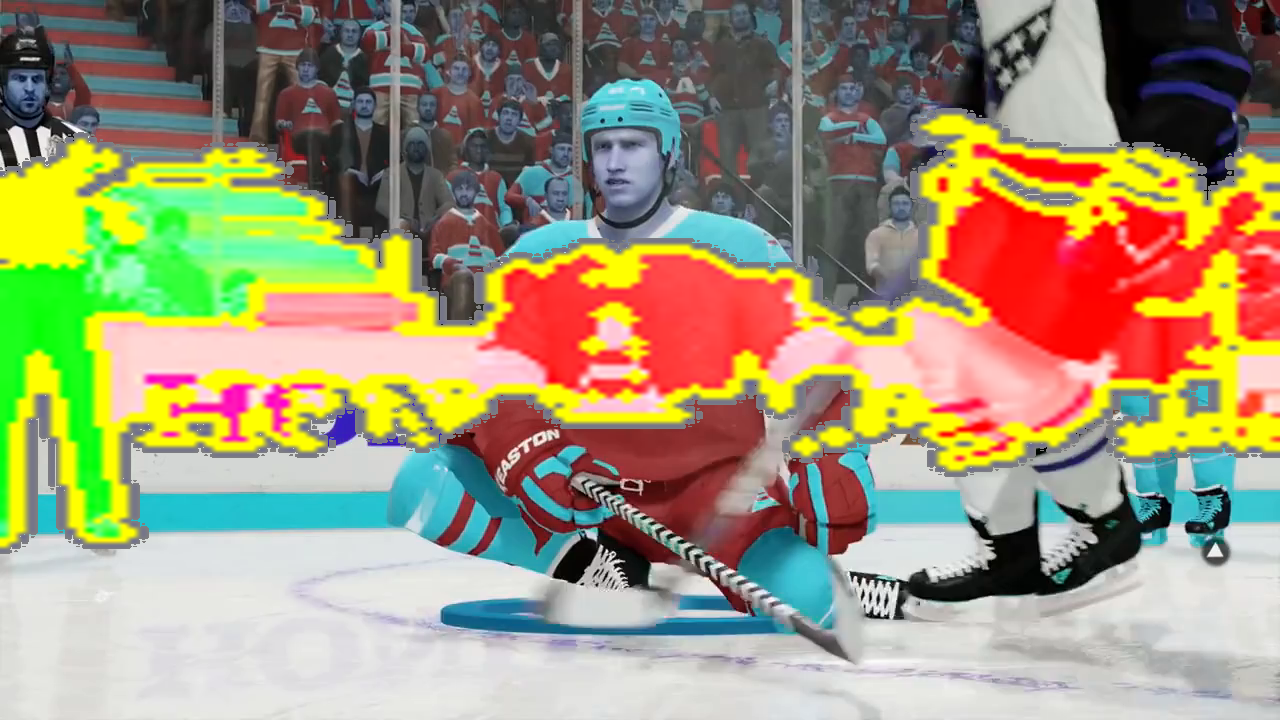}\hfill
    \caption{Mapping over the Sliding Goalie}\hfill
    \label{fig:slideGoalie}\hfill
    \end{subfigure}
    \begin{subfigure}{1\textwidth}
    \includegraphics[width=.48\textwidth]{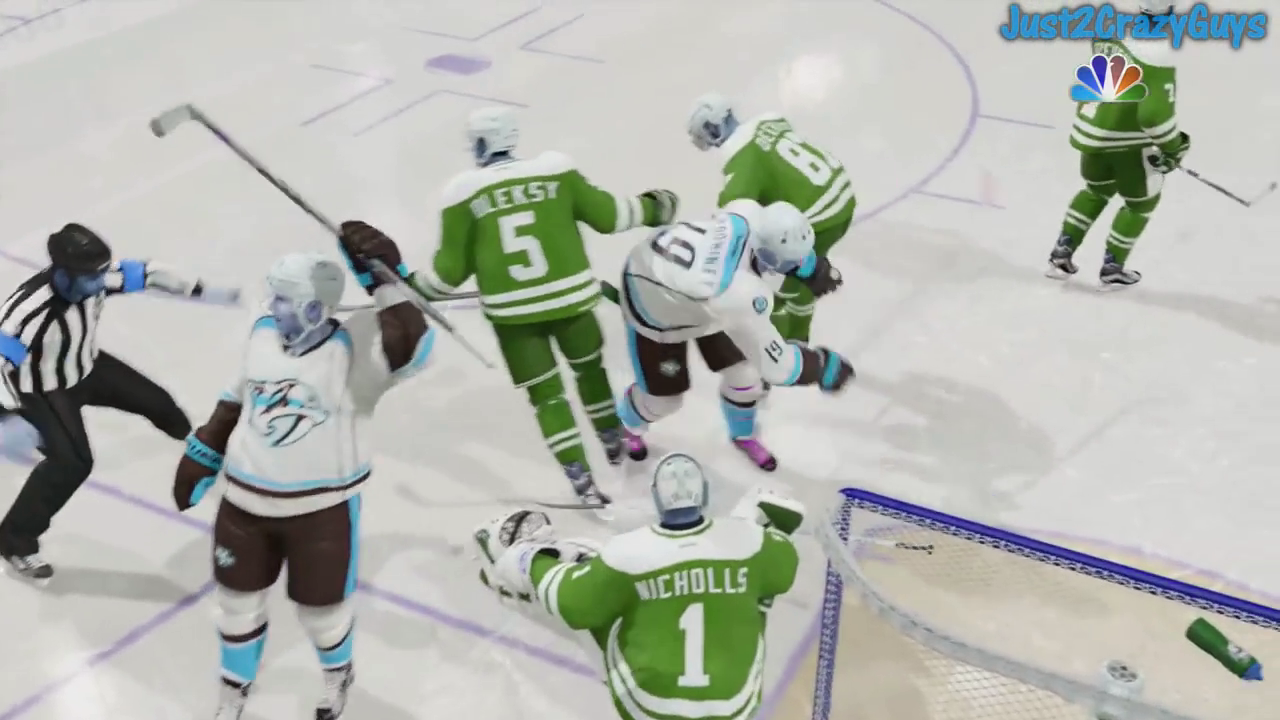}\hfill
    \includegraphics[width=.48\textwidth]{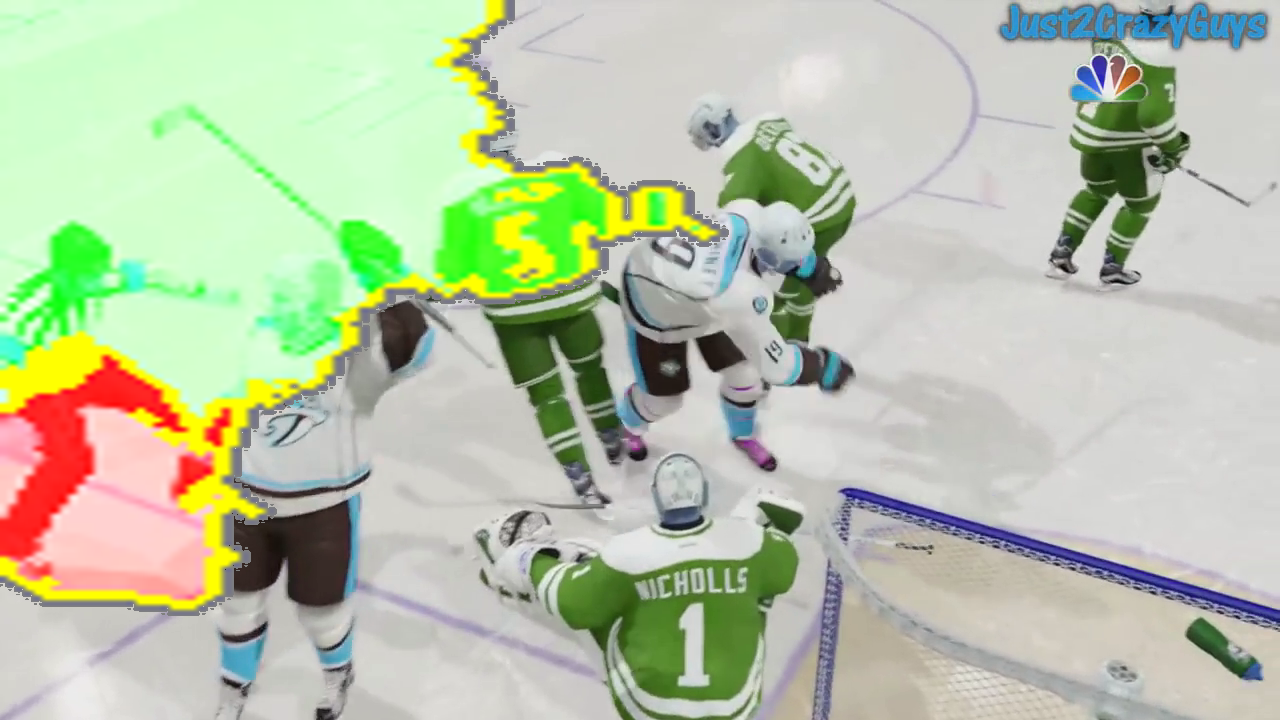}\hfill
    \caption{Mapping that favors closely gathered characters over the Sliding Goalie}\hfill
    \label{fig:deniedGoalie}
    \end{subfigure}
    \vspace*{-10mm}
    \caption{Mappings for the Sliding Goalie}
    \label{fig:goalieMap}
\end{figure}

\textbf{The Sliding Goalie:} For whatever reason, there were a noteworthy number of bugs in our dataset specifically involving players in hockey games (usually the goalie) being able to slide around the rink while sitting or laying down in unusual positions, as seen back in Figure~\ref{fig:exampleMap} or in ~\ref{fig:slideGoalie}. And for whatever reason, the model seemed to be able to identify those sliding goalies with a decent level of consistency. In fact, the most prominent example of the model missing a sliding goalie can be seen in Figure~\ref{fig:deniedGoalie}. In this example, the goalie is not covered by any pixel maps, but it could be because this image is also an example of closely gathered characters. Perhaps the Sliding Goalie has a lower level of importance in comparison.

\subsection{RQ5: What benefits are brought to developers using our tool in identifying buggy game play segments in videos over a manual analysis?}

To answer this question, we wanted to investigate if there was any benefit to using our tool for buggy segment detection besides high efficiency compared to manual analysis.  \observation{Our results show that participants on an average gave smaller buggy ratio ($0.60$) compared to the automated approach ($0.77$). 
Our results also show that participants on average had a higher start time ratio ($0.21$) when compared to the automated approach ($0.14$)}
The buggy ratio results indicate that manual analysis is more likely to miss important information for buggy segment detection. 
The start time ratio results mean that participants on average took longer to identify the first buggy segment (see Section 3.3 for a definition of buggy ratio and start time ratio).

%!TEX root = main.tex

\section{Discussion}
\label{sec:discussion}
In this section we discuss our results and their implications for researchers and practitioners.  
\subsection{For Researchers}
\textbf{Predictor Performance:} Our results indicate that our prediction models are capable of predicting whether a bug is presented in a video segment. When tested on our dataset from Section~\ref{sec:result_RQ1}, the top models returned F1 scores of 0.87 and 0.88. Even after creating smaller subset datasets at the genre level and at the game level in Sections~\ref{sec:result_RQ2} and ~\ref{sec:result_RQ3}, the top models still returned F1 scores of at least 0.70. The worst performance came from the models for \textit{Ark: Survival Evolved}, where the highest F1 score by any model across either clustering technique was 0.73. This could be due to the fact that, as the domain becomes more specific, the data might be more focused, but the number of data points decrease as well. This dataset only had 694 total video segments; if more video segments were added to this dataset in the future, this kind of evaluation might show improved results.  

Overall, however, the models produced promising performance results, significantly outperforming the prior state-of-the-art, GELID, across all datasets. One likely reason for this outperformance is likely due to the fact that GELID was designed to work specifically on videos that contained certain keywords in their transcripts~\cite{guglielmi2023using}. Because our video segment dataset was constructed without any restrictions on the content of the video transcripts, it is likely that GELID was unable to account for transcripts that fell outside the scope of what it was created for. 
For future work, researchers can look to devising techniques to even further improve the results of our classification framework. They might investigate the differences between the subsets of game videos, such as why sports games seemed to return better scores than action games and why \textit{FIFA 17} seemed to fare better than \textit{Ark: Survival Evolved}. Additionally, researchers can implement our automated approach in order to identify buggy segments of gameplay videos into their own research tasks on game bugs. Future work might examine the characteristics of these buggy segments or the types of bugs that appear in these segments.

\textbf{Superpixels:} The analysis of the superpixel maps was preliminary in nature. We wanted to investigate what kinds of objects the model was more likely to highlight and in what contexts. For example, the model seemed to have a tendency to highlight people, preferably people who are in groups. The most notable exception to this general observation is in Figure~\ref{fig:shapeSoccer} where the superpixels appeared to effectively ignore the players on the soccer field and instead ended following along with the various boundary lines. This could be explained by the difference in game view. The view in Figure~\ref{fig:shapeSoccer} is considerably further away from the people seen in the other images. Future researchers may choose to explore some of the nuances between different video images and what effect these nuances might have on the model's ability to correctly classify the presence of bugs. 

\textbf{User Study:} Our findings indicate that our automated bug identification models  tend to achieve better performance than manual analysis. For future researchers, they could implement our approach on different kinds of game videos to study if there exist differences in the distribution of buggy segments based on other attributes of the game videos. For instance, they could study how the length of the video, the game play platform, and game genres affect the attributes we extracted. Because we only performed our user study with five videos, there is still room for future work to examine some of these other characteristics when comparing model performance with user studies.

\subsection{For Practitioners}
\textbf{Predictor Performance:} 
Practitioners can use our bug prediction framework to help identify bugs in videos. 
Game developers wishing to apply our prediction framework to a narrow set of videos will likely still experience a high level of performance based on the results of our study.
The fact that the prediction results may or may not improve as the dataset becomes more specific may be particularly relevant to practitioners, since they are more likely to be interested in detecting bugs in games that they actively work on rather than over a range of different games. Practitioners may not be interested in manually designating clusters and buggy centroids, so the fact that the models still work with high performance even when using automatically generated clusters is certainly helpful.
Additionally, the results of our user study indicate that our framework provides benefits pertaining to accuracy and time saved when compared to manual analysis.  Overall, the tool we built could facilitate the whole process of issue tracking, while maintaining the high accuracy of prediction results.

\textbf{Superpixels:} Practitioners may be interested in running the superpixel pipeline on videos of their games since they are likely to have more domain knowledge about these games, and as such, may be better able to identify when a superpixel map is highlighting something significant.

%!TEX root = main.tex

\section{Threats to Validity}
\label{sec:threats}

Though we structured our study to eliminate the possible effects of random noise and avoid introducing bias, it is still possible that our mitigation strategies may not have been effective. This section reviews the threats to the validity of our study

The dataset used for the study contains game videos from YouTube as the single source. It is possible that game videos from other media platforms may have a different distribution in terms of buggy segments. Since all videos are collected from YouTube, our findings may be limited to game videos posted through YouTube. However, we believe that the collected dataset covers a wide range of video games which is sampled more than adequately for addressing this threat.

To fully understand the context of each video besides visual information, we also used the encoded transcript as a feature for training our classifiers. It is possible that the caption auto-generation tools on YouTube and Otter.ai could not address the context properly. Though the effectiveness of our classifier is limited by the current state-of-art technology of speech-to-text software, we believe that our labelling results are only likely to be improved as the technology evolves. To get deeper insight on the labelling results, we conducted multiple statistical tests to compare attributes between different game genres, as well as the user study and mining results. These tests would increase the chance of getting Type I error. To mitigate this threat, we did the Bonferroni correction~\cite{benjamini2001control}.

For the segmentation of videos, it is possible that the approach we used to split the videos might not have returned the best performance. We automatically split the videos based on the timestamps from the transcript, such as those presented in Figure~\ref{fig:captionExample}. We used this automated approach to minimize disruptive splits in the videos. While this approach did split phrases into multiple segments, it managed to avoid more awkward splits, such as splits in the middle of a word. Furthermore, the prior state-of-the-art from Guglielmi et al. also used the transcript timestamps in order to divide videos into segments~\cite{guglielmi2023using}. It is also possible that the timestamp and captions in the transcripts were inaccurate to each other (e.g., spoken language appearing in a different segment than what the transcript specified). We did not experience any inaccuracies when we labelled our segments, however.

From the user study result, it is always possible that the participants misinterpreted the context of the video and mistakenly perceived one buggy segment as there are no bugs included. To mitigate this threat, we conducted a pilot study with students who have some experience playing video games. The selected five videos for the user study were updated based on the findings of this pilot study. However, our findings through the user study may not be generalizable due to the small selected sample size, as there are only five videos. Since the user study is conducted to triangulate our mining result and the selected videos do cover a wide range of pattern on buggy segment distributions, we believe that this will not negatively impact our findings.
%!TEX root = main.tex

\section{Conclusion}
\label{sec:conclusion}  

In this paper, we introduced an automated approach based on machine learning to identify whether a segment of a gameplay video contains occurrences of bugs. The approach was evaluated using a dataset comprised of 4,412 video segments from 198 gameplay videos collected from YouTube. We trained and tested multiple machine learning models with these video segments, deriving features from the transcript text and the visual content. Testing our models returned a top F1 score of 0.88, indicating that this is a promising approach to identifying the presence of bugs in gameplay segments. Additionally, we found that training and testing models on video segments from the same genre of game still produced high F1 scores for both game genres evaluated (0.84 for action and 0.90 for sports). Performance was still high when training and testing models with video segments from a single game (0.73 for \textit{Ark: Survival Evolved} and 0.91 for \textit{FIFA 17}). In all cases, our approach significantly outperformed the prior state-of-the-art for classifying buggy segments of gameplay videos. Furthermore, we performed a superpixel analysis to identify possible patterns in the content of frame images that may hold importance to the classifier. Finally, we conducted a preliminary user study that indicated that our approach can identify buggy segments sooner and with greater accuracy than manual analysis.

For future work, we intend to 
expand the scope of the genres and games used to train the models. Evaluation of models trained on additional video sets may reveal aspects of these videos that have an association with bugs. We also plan to explore the superpixel pipeline further.
\section{Data Availability}
The research artifacts for this study are available publicly at the companion website~\cite{suppMaterialM}. We intend to make the data and our analysis scripts publicly available upon publication of the paper.

\bibliographystyle{IEEEtranS}
\bibliography{bib}

\end{document}